\documentclass[amsmath, amssymb, aps, prl, reprint, floatfix, superscriptaddress]{revtex4-2}

\usepackage[T1]{fontenc}
\usepackage[utf8]{inputenc}
\usepackage{xcolor}
\definecolor{linkscolor}{RGB}{10,55,130}
\usepackage[pdftex, colorlinks=true, linkcolor=linkscolor, citecolor=linkscolor, urlcolor=linkscolor, breaklinks=true]{hyperref}
\usepackage[english]{babel}
\usepackage{braket}
\usepackage{graphicx}
\usepackage{dcolumn}
\usepackage{bm}
\usepackage{comment}
 \usepackage[normalem]{ulem}
\newcommand{\im}{\mathrm{i}}
\newcommand{\dif}{\mathrm{d}}

\newcommand{\el}{\mathrm{e}}
\newcommand{\DOS}{\mathcal{N}_{\mathrm{2D}}}

\definecolor{green2}{rgb}{0., 0.5, 0.}
\definecolor{blue2}{rgb}{0., 0.35, 1}
\definecolor{brown}{rgb}{0.7, 0.15, 0}
\definecolor{red}{rgb}{1, 0.1, 0.1}

\begin{document}

\title{Polariton-Assisted Inelastic Tunneling through a Quantum Well}

\author{Théophile Seck}
\affiliation{University of Strasbourg and CNRS, CESQ and ISIS (UMR 7006), 67000 Strasbourg, France}
\affiliation{Université de Strasbourg, CNRS, Institut de Physique et Chimie des Matériaux de Strasbourg, UMR 7504, F-67000 Strasbourg, France}
\author{Yanko Todorov}
\affiliation{Laboratoire de Physique et d'Etude des Matériaux, LPEM, UMR 8213, ESPCI Paris, Université, PSL, CNRS, Sorbonne Université, F-75005 Paris, France}
\author{Guido Pupillo}
\affiliation{University of Strasbourg and CNRS, CESQ and ISIS (UMR 7006), 67000 Strasbourg, France}
\author{David Hagenmüller}
\email{david.hagenmuller@ipcms.unistra.fr}
\affiliation{Université de Strasbourg, CNRS, Institut de Physique et Chimie des Matériaux de Strasbourg, UMR 7504, F-67000 Strasbourg, France}

\date{\today}

\begin{abstract}
We investigate electronic transport through a doped quantum well strongly interacting with a photonic mode confined in a double-metal cavity. Using a nonequilibrium Green's function formalism, we derive compact expressions for the current valid for arbitrary collective light--matter coupling strengths exceeding the relevant loss rates. We show that cavity polaritons leave observable transport signatures when the carrier injection rate is smaller than the cavity-induced electronic broadening. In this regime, the current--voltage characteristics exhibit inelastic sidebands associated with resonant and anti-resonant polariton emission, which are strongly enhanced under resonant illumination. Our results provide a realistic route to detecting cavity-induced modifications of charge transport in semiconductor heterostructures.
\end{abstract}

\maketitle

\begin{figure}[h!]
    \centering    
    \includegraphics[width=0.99\columnwidth]{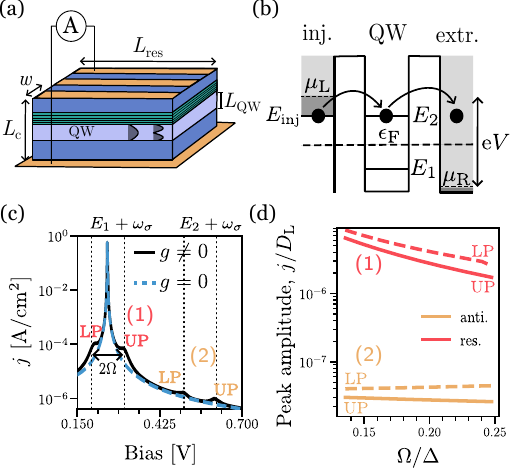}
    \caption{\textbf{Polariton-assisted inelastic tunneling in the dark.}
(a) A doped quantum well (QW) with electron density $\rho$ is embedded in an array of mid-infrared wire cavities. The fundamental mode of each cavity (quality factor $Q=\omega_\mathrm{cav}/\kappa$, where $\kappa$ is the cavity decay rate) couples resonantly to $N=\rho\,wL_\mathrm{res}$ electrons through the intersubband transition $\Delta=E_2-E_1=150\,\mathrm{meV}$, with single-electron coupling strength $g$. Electrons tunnel from an injector miniband of energy $E_{\mathrm{inj}}$ and bandwidth $D_\mathrm{L}=\mu_\mathrm{L}-E_{\mathrm{inj}}$. 
(b) Resonant tunnelling into the second subband ($E_{\mathrm{inj}}=E_2$).
(c) Current--voltage characteristics, where $j = J/w L_\mathrm{res}$ denotes the current density. For $g=0$ (blue dashed), only the resonant peak of panel (b) is present. For $g\neq0$ (black), additional inelastic peaks appear at $E_1+\omega_\sigma$ and $E_2+\omega_\sigma$, arising from resonant and anti-resonant polariton emission, respectively. Here $\omega_\sigma$ ($\sigma=\mathrm{LP},\mathrm{UP}$) are the lower- and upper -polariton frequencies, split by the vacuum Rabi energy $2\Omega \simeq 0.25\Delta$. The Fermi level $\epsilon_{\mathrm{F}}$ lies near mid-gap between the subbands ($\rho \simeq 7\times10^{11}\,\mathrm{cm^{-2}}$).
(d) Inelastic peak amplitudes versus collective coupling $\Omega=g\sqrt{N}$ at fixed $g$, with the Fermi level tuned between the two subbands. Parameters: $\Gamma=0.01\,\mathrm{meV}$, $Q = 10$, $w=L_\mathrm{c}=100\,\mathrm{nm}$, $L_\mathrm{QW}=10.6\,\mathrm{nm}$, $L_\mathrm{res} = 1.2\, \mathrm{\mu m}$, corresponding to $g/\Delta\simeq0.006$.}
    \label{figure1}
\end{figure}

The strong coupling regime is reached when the light--matter coupling strength exceeds the relevant loss rates. In semiconductor heterostructures, it has been extensively studied through optoelectronic signatures such as electrical control of polaritons~\cite{Anappara2006,limbacher2020resonant}, electroluminescence~\cite{Sapienza_Vasanelli_2008,De_Liberato_Ciuti_2009}, and photodetection~\cite{Vigneron2019}. Beyond these optical manifestations~\cite{unterrainer1999photon}, strong coupling can also modify material properties in the dark~\cite{RevModPhys.85.299,garcia-vidal_manipulating_2021,Schlawin2022}, and its impact on charge transport has attracted considerable attention.

Several examples of cavity-modified charge transport have been reported under strong light--matter coupling, ranging from conductivity enhancement in organic systems~\cite{orgiu_conductivity_2015,hagenmuller_2017,hagenmuller2018,Nagarajan2020,Kumar2024} to cavity control of transport in quantum Hall devices~\cite{ciuti_cavity-mediated_2021,ArwasCiuti2023Quantum,PhysRevLett.131.196602,Rokaj2022,Appugliese2022barnesreakdown,Enkner2024Testing,Bacciconi2025_frac,Enkner2025Tunable,Mattiotti2026EnhancedLocalization}. Equilibrium transport through a quantum well (QW) has also been predicted to be modified by the cavity in the ultrastrong-coupling regime~\cite{Naudet-Baulieu2019}. However, the corresponding conductivity calculation relies on phenomenological polariton lifetimes, making the preservation of causality unclear. Linear-response calculations of the optical conductivity suggested that cavity-induced equilibrium transport features are governed by the typically weak single-electron coupling strength and should therefore be negligible~\cite{Amelio_Korosec_Carusotto_Mazza_2021}.

In this Letter, we demonstrate that the conclusion drawn from equilibrium linear-response theory does not extend to the non-equilibrium transport regime: cavity polaritons leave measurable fingerprints in the charge current through a doped QW embedded in a double-metal cavity [Fig.~\ref{figure1}(a),(b)]. Using a microscopic nonequilibrium Green's function (NEGF) framework, which inherently preserves causality, we derive compact analytical expressions for the current that remain valid for arbitrary collective coupling strengths between the first intersubband (ISB) transition and the fundamental transverse magnetic (TM) cavity mode, provided they exceed the relevant loss rates. We show that cavity effects emerge in the slow-injection regime, where the carrier injection rate is smaller than the cavity-induced broadening of the electronic states. In this regime, the current--voltage characteristics develop inelastic sidebands arising from resonant and anti-resonant polaritonic emission processes [Fig.~\ref{figure1}(c)]. While the sideband energies are determined by the collective light--matter coupling through the polariton frequencies, their amplitudes grow with the single-electron coupling strength and the cavity quality factor [Fig.~\ref{figure1}(d)]. Finally, including a realistic electronic lifetime limited by phonon and impurity scattering, we show that resonant illumination strongly enhances the inelastic transport channels, providing a realistic route toward the experimental observation of cavity-induced transport signatures in semiconductor heterostructures.

More broadly, our work suggests that signatures commonly attributed to electrical injection into polaritonic states in ISB devices~\cite{Bajoni_Semenova_Lemaitre_Bouchoule_Wertz_Senellart_Bloch_2008,Sapienza_Vasanelli_2008,Jouy_Vasanelli_Todorov_Sapienza_Colombelli_Gennser_Sirtori_2010} and cavity-coupled organic semiconductors~\cite{orgiu_conductivity_2015,Nagarajan2020,Kumar2024} may instead originate from polariton-assisted inelastic tunneling processes.

We consider a two subband QW containing $N$ electrons, coupled to the fundamental TM mode of a wire cavity, whose resonance is determined solely by the resonator length $L_\mathrm{res}$~\cite{Feuillet-Palma:12} [Fig.~\ref{figure1}(a)]. Such cavities combine a small electrical area with a small mode volume, the latter enabling large single-electron light--matter coupling strengths. The cavity mode is resonant with the ISB transition of frequency $\Delta$, and the system is connected to source and drain contacts. We assume zero temperature and neglect Coulomb interactions, which are expected primarily to renormalize the ISB transition frequency~\cite{Todorov2012}. The Hamiltonian $H = H_\mathrm{elec} + H_\mathrm{cav} + H_\mathrm{bath} + H_\mathrm{int}$ consists of the electronic part $H_\mathrm{elec}=H_\mathrm{QW}+H_\mathrm{leads}+H_\mathrm{tun}$, with $H_\mathrm{QW}=\sum_{\mathbf{k}, n}\epsilon_n(k) \, \hat c^\dagger_{n\mathbf{k}} \hat c^{\phantom{\dagger}}_{n\mathbf{k}}$, with $\hat c_{n\mathbf{k}}$ annihilating an electron of in-plane wavevector $\mathbf{k}$ in subband $n$ and energy
$\epsilon_n(k)=\hbar^2k^2/(2m^*)+E_n$, where $k\equiv|\mathbf{k}|$, $m^*$ is the electron effective mass, and $E_n$ is the bottom of subband $n$. The contact Hamiltonian $H_\mathrm{leads}$ describes a narrow injector miniband of width $4t$ centered at $E_\mathrm{inj}$ and shifted by the applied bias according to $E_\mathrm{inj}(V)=E_\mathrm{inj}(0)+eV/2$, while the extractor is modeled as a three-dimensional (3D) electron gas. The tunneling Hamiltonian $H_\mathrm{tun}$ describes electron transfer between the QW and contacts. We assume identical injection and extraction rates $\Gamma$. The cavity Hamiltonian is $H_\mathrm{cav}=\hbar\omega_\mathrm{cav}\hat a^\dagger\hat a$, where $\hat a$ is the bosonic annihilation operator. Cavity losses at rate $\kappa$ are described by $H_\mathrm{bath}$ (see Supplemental Material~\cite{Note1}). The light--matter interaction couples the cavity mode to the first ISB transition and reads
\begin{equation}
    \label{equation1}
    H_\mathrm{int} = \hbar g \sum_{\mathbf{k}} \left(\hat c^\dagger_{2\mathbf{k}} \hat c^{\phantom{\dagger}}_{1\mathbf{k}} + \mathrm{h.c.}\right) \left(\hat a + \hat a^\dagger\right) + \hbar D \left(\hat a + \hat a^\dagger\right)^2,
\end{equation}
where $g$ is the \textit{single-electron} coupling strength, determined by the cavity mode volume $V = w\, L_\mathrm{c}\, L_\mathrm{res}$, and $D=g^2 N/\Delta$. Since the photon in-plane momentum is negligible compared to the Fermi wavevector $k_{\mathrm F}$, it is neglected throughout. The light--matter interaction in Eq.~\eqref{equation1} contains both resonant and anti-resonant emission processes, $\propto \hat a^\dagger \hat c^\dagger_{1\mathbf{k}} \hat c^{\phantom{\dagger}}_{2\mathbf{k}}$ and $\propto \hat a^\dagger \hat c^{\dagger}_{2\mathbf{k}} \hat c^{\phantom{\dagger}}_{1\mathbf{k}}$, respectively, leading to distinct transport signatures discussed below. The collective eigenmodes of $H_\mathrm{QW}+H_\mathrm{int}+H_\mathrm{cav}$ are lower and upper polaritons at energies $\omega_\pm$ separated by the vacuum Rabi splitting $2\Omega$, where $\Omega=g\sqrt{N}$ is the collective coupling strength~\cite{Ciuti_Bastard_Carusotto_2005}.

\begin{figure*}[htbp]
    \centering
    \includegraphics[width=\textwidth]{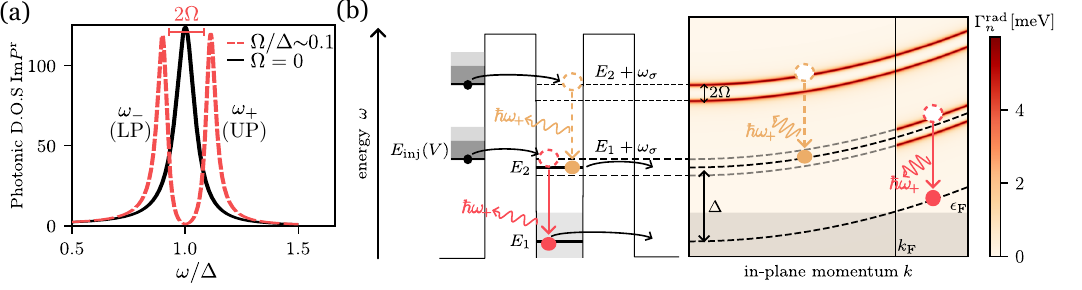}
    \caption{\textbf{Cavity-induced radiative processes.}
(a) Photonic density of states, $\mathrm{Im}\,P^\mathrm{r}$, without (black) and with (dashed red) light--matter coupling.
(b) Equilibrium radiative broadening function (right) and associated inelastic tunneling channels (left). Red (dashed orange) arrows denote resonant (anti-resonant) light--matter processes. Parameters are as in Fig.~\ref{figure1}.}
    \label{figure2}
\end{figure*}

To identify the signatures of these collective polariton modes in charge transport through the QW, we evaluate the current from the injector miniband within the NEGF formalism in the Meir--Wingreen framework~\cite{Wingreen1988,Wingreen1989,Meir1992,HaugJauho2008}. The steady-state current can be expressed as~\cite{Note1}
\begin{align}
    \label{equation2}
    J &= J_0 \int \mathrm{d}k \, \mathrm{d}\omega \, F_{k}(\omega, V)\, T(k, \omega)\left[f_\mathrm{L} - f_\mathrm{R}\right],
\end{align}
where $J_{0}=2\el\Gamma \DOS t/h$, and $\DOS$ denotes the 2D density of states. The lead occupations are described by the Fermi functions
$f_\eta(\omega,V)=\Theta\left[\mu_\eta(V)-\omega\right]$, with $\Theta$ the Heaviside function and $\mu_\eta(V)=\mu_\eta(0)\pm \el V/2$ the chemical potential of lead $\eta=\mathrm{L,R}$. At equilibrium ($V=0$), $\mu_{\mathrm{L}}=\mu_{\mathrm{R}}=\epsilon_{\mathrm{F}}$, with $\epsilon_{\mathrm{F}}=\hbar^2 k_{\mathrm{F}}^2/(2m^*)$ the Fermi energy and $k_{\mathrm{F}}$ the Fermi wavevector. Here $F_k(\omega,V)$ encodes the injector geometry and acts as a supply function, determining the number of states available for tunneling. It imposes the injection condition $\omega \geq E_\mathrm{inj}(V)$ [Fig.~\ref{figure1}(b)]. The injector bandwidth is defined as $D_\mathrm{L}\equiv\mu_\mathrm{L}(0)-E_\mathrm{inj}(0)$. The tunneling channels in the QW are encoded in the transmission function $T(k,\omega)=-\sum_n \operatorname{Im} G_n^\mathrm{r}(k,\omega)$, where the retarded Green's function satisfies the Dyson equation
$G_n^\mathrm{r}(k,\omega)=\left[\epsilon_n(k)-\omega-\Sigma_n^\mathrm{r}(k,\omega)\right]^{-1}$. The retarded self-energy $\Sigma_n^{\mathrm{r}}(k,\omega)$ includes both radiative and tunneling contributions, originating from the light--matter interaction [Eq.~\eqref{equation1}] and from the QW--contact coupling, respectively. It can be decomposed as $\Sigma_n^{\mathrm{r}}(k,\omega) = \Lambda_n(k,\omega) + \im\Gamma_n(k,\omega)/2$. Assuming Markovian contacts, the real part $\Lambda_n$ arises solely from the light--matter interaction and accounts for a small renormalization of the QW energy levels~\cite{Note1}. Its effect is therefore not important in the present context. The tunneling self-energy is purely imaginary and broadens the QW energy levels by an amount $\propto \Gamma$. The total broadening is $\Gamma_n=\Gamma_n^{\mathrm{rad}}+2\Gamma$, with $\Gamma_n^{\mathrm{rad}}$ originating from the light--matter interaction. In the limits $t\ll\Gamma_n$ and $\el V>D_\mathrm{L}$, Eq.~\eqref{equation2} simplifies to
\begin{equation}
    \label{equation3}
    J = J_0 \sum_n \int_{D_\mathrm{L}} \frac{\mathrm{d}\omega \, \Gamma_n/2}{\left(E_\mathrm{inj}(V) - E_n - \Lambda_n\right)^2 + \left(\Gamma_n/2\right)^2},
\end{equation}
Here, the in-plane momentum $k$ entering $\Gamma_n(k,\omega)$ and $\Lambda_n(k,\omega)$ is fixed by momentum conservation and can therefore be expressed as
$k(V)=\sqrt{2m[\omega-E_\mathrm{inj}(V)]}/\hbar$. 

In the absence of light--matter coupling ($g=0$), $\Gamma_n=2\Gamma$ and $\Lambda_n=0$, and Eq.~\eqref{equation3} reduces to a sum of Lorentzian resonances centered at $E_\mathrm{inj}(V)=E_n$, corresponding to alignment of the injector with the bare subband levels [blue curve in Fig.~\ref{figure1}(c)]. In this regime, carriers tunnel elastically into quasiparticle states of the QW, yielding a peak current $J_{0}D_{\mathrm{L}}/\Gamma$.

In the presence of light--matter coupling ($g\neq 0$), electronic states are dressed by the cavity fluctuations, which provide a radiative contribution $\Gamma^\mathrm{rad}_n$ to the broadening of QW levels. The dark current develops additional resonances in its current--voltage characteristics [black curve in Fig.~\ref{figure1}(c)]. Two appear on either side of the dominant quasiparticle resonance and stem from the resonant terms in Eq.~\eqref{equation1}, while the two weaker high-energy resonances originate from the anti-resonant terms. The two resonances in each pair are separated by the collective Rabi splitting $2\Omega$. In Fig.~\ref{figure1}(d), we investigated the dependence of those current resonances on $\Omega$, which can be tuned by varying the electron density $\rho$ in the QW while keeping $g$ fixed. The resonances associated with the resonant terms are strongly suppressed as $\Omega$ increases, whereas those originating from the anti-resonant terms remain nearly unchanged, as explained below.

To understand the microscopic transport mechanisms underlying Fig.~\ref{figure1}(c), we analyze the radiative broadening entering Eq.~\eqref{equation3},
\begin{equation}
\label{rad_SE}
\Gamma^\mathrm{rad}_n (k, \omega) = g^2 (1-\delta_{n,n'}) \int \dif \omega' G^{>}_{n'}(k, \omega' + \omega) P^{<}(\omega'),
\end{equation}
which is second order in the single-electron coupling strength $g$~\cite{Amelio_Korosec_Carusotto_Mazza_2021}. Here, $G^{>}_{n}(k,\omega)$ is the ``greater'' Green's function describing the unoccupied spectral weight associated with the state $(k,\omega)$ in subband $n$, and $P^{<}(\omega) = -2 \im \Theta(-\omega) \operatorname{Im}{P^{\mathrm{r}}(\omega)}$. The retarded polaritonic propagator reads~\cite{Note1}
\begin{equation}
\label{eq:polaritonic_propagator}
  P^\mathrm{r}(\omega) = \frac{2 \Delta(\omega^2 - \Delta^2)}{(\omega^2 - \omega_+^2)(\omega^2 - \omega_-^2) + 2\im \kappa \, \mathrm{sgn}(\omega)\Delta(\omega^2 - \Delta^2)}.
\end{equation}
In the strong-coupling regime $\Omega \gg \kappa$, collective effects are encoded in this propagator, whose poles are located at the polariton frequencies $\omega_{\pm}$.
If the Fermi level lies in between the two subbands, as we consider in this work, those are given by $\omega_\pm = \sqrt{\Delta^2 + \Omega^2} \pm \Omega$. The cavity photon spectral function $\operatorname{Im} P^\mathrm{r} (\omega)$ featuring two polariton peaks centered at $\omega_{\pm}$ is shown in Fig.~\ref{figure2}(a). 

At \textit{equilibrium}, Eq.~\eqref{rad_SE} can be simplified by replacing the interacting Green's function $G^{>}_{n}$ with its noninteracting counterpart, whose form is determined by the occupations of the QW subbands~\cite{Note1}. This approximation provides a transparent interpretation of the broadening function shown in the right panel of Fig.~\ref{figure2}(b). We find that two families of band replicas contribute to the radiative broadening, as commonly encountered in phonon-assisted processes~\cite{Wingreen1988,Wingreen1989,JonsonPRB1989,Anda_Flores_1991,Davies1993,Hyldgaard_Hershfield_Davies_Wilkins_1994,Galperin_Ratner_Nitzan_2004}. The two replicas at $\epsilon_1(k)+\omega_\pm$ contribute to $\Gamma^\mathrm{rad}_2(k,\omega)$ and originate from the decay of an electron injected into the second subband into an available state of the first subband at energy $\epsilon_1(k)$ by emission of a polariton of energy $\omega_{\pm}$. They therefore occur only for $k>k_\mathrm{F}$ and stem from the resonant terms $\propto \hat a^\dagger \hat c^\dagger_{1\mathbf{k}}\hat c_{2\mathbf{k}}$. Conversely, the replicas at $\epsilon_2(k)+\omega_\pm$ contribute to $\Gamma^\mathrm{rad}_1(k,\omega)$ and arise from the scattering of an electron injected into the first subband into the second subband accompanied by polariton emission. Since the final state is always empty, they exist for all $k$ and originate from the anti-resonant terms $\propto \hat a^\dagger \hat c^{\dagger}_{2\mathbf{k}}\hat c_{1\mathbf{k}}$.

\begin{figure*}[htbp]
    \centering
    \includegraphics[width=\textwidth]{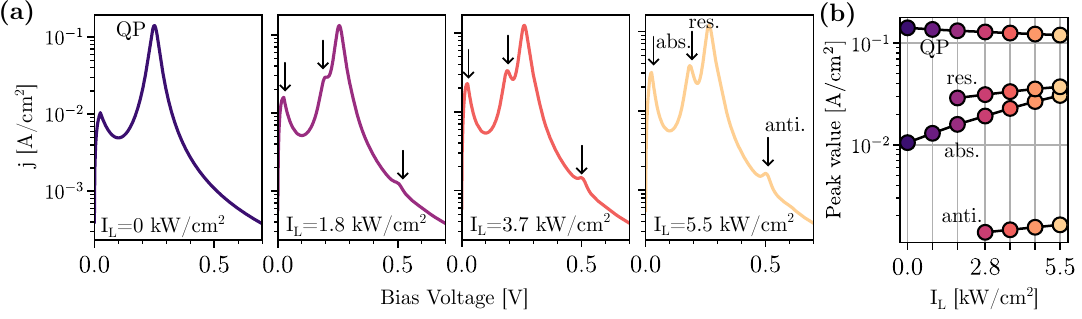}
    \caption{\textbf{Transport under illumination.}
(a) Current--voltage characteristics under resonant lower-polariton excitation. Arrows indicate the inelastic sidebands. Here $I_\mathrm{L}$ denotes the incident laser intensity.
(b) Peak current of the lower-polariton sidebands as a function of laser intensity. Parameters are the same as in (a), with $\gamma_\mathrm{el}=10\,\mathrm{meV}$~\cite{Ferreira_Bastard_1989}, corresponding to $\gamma_\mathrm{el}/(2g^2/\kappa)\sim10^2$, and a sample area (beam waist) of $2500 \,\mathrm{\mu m}^2$. (abs.): absorption, (res.): resonant, (anti.): antiresonant.} 
\label{figure3}
\end{figure*}

To compute the \textit{nonequilibrium current}, we write
$G^{>}_{n}(k,\omega)=2\im \pi \left[1-f_\mathrm{QW}(\omega, V)\right]\operatorname{Im}{G^{\mathrm{r}}_{n'}(k,\omega)}$ and substitute it into Eq.~\eqref{rad_SE}. We further adopt the pseudo-equilibrium approximation~\cite{Jauho1994} $f_\mathrm{QW}(\omega,V) = [f_\mathrm{L}(V)F_k(\omega , V)+f_\mathrm{R}(V)]/\left[1+F_k(\omega , V)\right]$, where the supply function $F_k(\omega,V)$ restricts injector contributions to states above the injection energy. This partially self-consistent scheme~\cite{Note2} determines the QW population from injection and extraction. In particular, unlike at equilibrium, the resonant process [red arrow in Fig.~\ref{figure2}(b)] remains allowed below $k_\mathrm{F}$ whenever $f_\mathrm{QW}=0$, i.e., when extraction into the right lead is possible. The only constraint is the injection window, $E_\mathrm{inj}(V)<\omega<\mu_\mathrm{L}(V)$, which imposes a maximum wavevector $k_\mathrm{max}=\sqrt{2m^*D_\mathrm{L}}/\hbar$. Equation~\eqref{equation3} is then solved numerically, with $\Lambda_n$ obtained from $\Gamma_n$ via the Kramers--Kronig relations, ensuring a causal self-energy and yielding the current--voltage characteristics shown in Fig.~\ref{figure1}(c).

A key advantage of the NEGF formalism is that it yields compact expressions for the self-energies and current within a controlled approximation scheme. In the strong-coupling regime, a closed-form expression can be obtained within the polariton-pole approximation, whereby the retarded propagator $P^\mathrm r$ is replaced by the sum of its residues at the polariton poles~\cite{Note1}. This gives
\begin{align}
\label{equation6}
&\Gamma^{\mathrm{rad}}_n (k, \omega) = 4\pi g^2 (1-\delta_{n,n'} ) \left[1-f_\mathrm{QW}(\epsilon_{n'}(k), V)\right]  \nonumber \\
& \times \sum_{\sigma=\pm} Z_{\sigma}\frac{Z_{\sigma}\kappa}{(\omega - \epsilon_{n'}(k) - \omega_{\sigma})^2 + (Z_{\sigma}\kappa)^2},
\end{align}
where $Z_{\sigma}$ is a dimensionless factor related to the photonic weight of polariton branch $\sigma$. Since both Eq.~\eqref{equation6} and the corresponding energy shift are proportional to $1-f_\mathrm{QW}[\epsilon_{n'}(k),V]$, in-plane momentum conservation, $k\rightarrow k(V)$, allows the current to be expressed solely in terms of the bias voltage. For $e V>D_\mathrm{L}$, the frequency integral in Eq.~\eqref{equation3} yields a factor $D_\mathrm{L}$, and the current separates into elastic and inelastic contributions~\cite{Note1,Wingreen1989,Davies1993,HYLDGAARD1994}, $J=J_\mathrm{el}+J_\mathrm{inel}$, with
\begin{align}
    \label{equation7}
    J_\mathrm{el}&= J_0 D_{\mathrm{L}} \sum_n \frac{\Gamma}{\left( E_\mathrm{inj} - E_n \right)^2 + \left(\Gamma_n/2\right)^2} \\
    \label{equation8}
    J_\mathrm{inel} &= J_0 D_{\mathrm{L}} \sum_{n} \frac{ \Gamma^{\mathrm{rad}}_{n}}{\left( E_\mathrm{inj} - E_n \right)^2 + \left(\Gamma_n/2\right)^2},
\end{align}
Equations~\eqref{equation7} and \eqref{equation8} are a central result of this work. Here, $\Gamma^{\mathrm{rad}}_n(V)$ is obtained from Eq.~\eqref{equation6} by enforcing momentum conservation, $\epsilon_{n'}(k)\rightarrow \omega-E_\mathrm{inj}(V)+E_{n'}$. $\Lambda_n \sim g^2/\kappa \ll \Delta$, and is therefore neglected in the denominators. Equation~\eqref{equation7} then yields a single quasiparticle resonance at $E_\mathrm{inj}(V)=E_n$, whose peak amplitude is reduced by a factor $(\Gamma/\Gamma_n)^2$ relative to the $g=0$ case [Fig.~\ref{figure1}(c)], as previously found for electron--phonon interactions~\cite{Davies1993}. The cavity-induced features instead arise from Eq.~\eqref{equation8} and occur when $\Gamma^{\mathrm{rad}}_{n}(V)$ is maximal, i.e., at $E_\mathrm{inj}(V)=E_{n'}+\omega_\sigma$. Because this condition does not minimize the denominator of Eq.~\eqref{equation8}, these features are not quasiparticle resonances but inelastic satellite peaks associated with polariton emission, with a width controlled by $\kappa$ and $g$. They become visible only when they exceed the tail of the quasiparticle resonance, namely for $\Gamma<\Gamma^{\mathrm{rad}}_{n} \sim 2 g^2/\kappa$. Thus, observing cavity-induced transport signatures requires the slow-injection regime, where the electron dwell time in the QW exceeds the characteristic polariton emission time.

Having established that the cavity-induced inelastic peaks are primarily controlled by the single-electron coupling strength $g$, we now elucidate the role of collective effects. Neglecting cavity-induced corrections in the denominator of Eq.~\eqref{equation8}, the peak amplitude scales as $J_\mathrm{inel} \sim \frac{J_0 D_\mathrm{L} g^2}{\kappa(\Delta \pm \omega_\sigma)^2}$, where the minus (plus) sign corresponds to resonant (anti-resonant) processes. Collective effects enter only through the polariton energies $\omega_\sigma$. Outside the ultrastrong-coupling regime, $\omega_\sigma \simeq \Delta + \sigma \Omega$, yielding $J_\mathrm{inel} \sim  \frac{J_0 D_\mathrm{L} g^2}{\kappa\Omega^2}$ for resonant processes and $J_\mathrm{inel} \sim  \frac{J_0 D_\mathrm{L} g^2}{\kappa(2\Delta + \sigma \Omega)^2}$ for anti-resonant ones. This explains Fig.~\ref{figure1}(d): the resonant peaks are suppressed as $\Omega$ increases (for fixed $g$), whereas the anti-resonant peaks remain nearly unchanged for $\Omega\ll\Delta$. Appreciable modifications of the latter therefore emerge only in the collective deep-strong-coupling regime, $\Omega\sim\Delta$. The satellite peaks are thus maximized in an intermediate regime where the collective coupling exceeds the loss rates while remaining sufficiently small that the sidebands lie close to a quasiparticle resonance, minimizing the denominator of Eq.~\eqref{equation8}.

In a realistic setup, however, the electronic lifetime is intrinsically limited by optical-phonon and impurity scattering, corresponding to a broadening of $\gamma_\mathrm{el}\sim10\,\mathrm{meV}$ in a GaAs QW~\cite{Ferreira_Bastard_1989}. This broadening is therefore expected to reduce the visibility of the inelastic current peaks under dark conditions. Importantly, the radiative self-energy scales with the cavity photon population and is therefore expected to be strongly enhanced under coherent illumination~\cite{Henrickson_2002,Aeberhard_2012}. We therefore investigate the fate of the cavity-induced inelastic current peaks under coherent driving, considering a value of $\gamma_\mathrm{el}$ for which no inelastic features are visible in the dark current [Fig.~\ref{figure3}(a)]. We assume that a laser of intensity $I_\mathrm{L}$ excites the lower polariton at energy $\omega_-$, thereby populating the cavity with a coherent state~\cite{Note1}. Increasing the laser intensity progressively enhances the lower-polariton inelastic sidebands associated with both resonant and anti-resonant emission processes, while suppressing the elastic quasiparticle resonance. This behavior is the hallmark of stimulated-emission-assisted tunneling, akin to photon-assisted transport~\cite{Tien_Gordon_1963,Guimaraes_Keay_Kaminski_Allen_Hopkins_Gossard_Florez_Harbison_1993,Foden_Whittaker_1998, drexler1995photon, Platero_Aguado_2004,Kawano_Fuse_Toyokawa_Uchida_Ishibashi_2008} that has been well described in QWs \cite{unterrainer1999photon}. Simultaneously, a low-bias resonance emerges from transitions between the two subbands assisted by lower-polariton absorption. The corresponding peak amplitudes are shown in Fig.~\ref{figure3}(b). At sufficiently high intensities, additional resonances and transport gaps appear as the drive-induced energy shift becomes comparable to the ISB gap, substantially modifying the quasiparticle resonance itself~\cite{Espinosa-Ortega_Kyriienko_Kibis_Shelykh_2014,Morina_Kibis_Pervishko_Shelykh_2015,Valmorra_et_al_2021}.

In conclusion, we have investigated electron transport through a QW coupled to a cavity. We show that, in the slow-injection regime, cavity coupling opens inelastic tunneling channels associated with polariton emission, yielding distinct transport signatures that are strongly enhanced under illumination. More broadly, our work establishes a new link between strong light--matter coupling and inelastic tunneling spectroscopy, suggesting that the large enhancement of charge transport reported in cavity-coupled organic semiconductors~\cite{orgiu_conductivity_2015,Nagarajan2020,Kumar2024} may originate from polariton-assisted inelastic tunneling processes. Our results also open the way to the study of electroluminescence and photocurrent generation in polaritonic devices~\cite{Lagree_2022,Pisani_2023}, as well as polariton-assisted transport in the dynamical Coulomb blockade regime~\cite{Devoret_1990,Girvin_1990}, potentially realizable using THz metamaterials coupled to semiconductor heterostructures~\cite{Iqbal_Todorov_Mora_2024}.

\textit{Acknowledgements} --- This work of the Interdisciplinary Thematic Institute QMat, as part of the ITI 2021-2028 program of the University of Strasbourg, CNRS, and Inserm, was supported by IdEx Unistra (Project No. ANR 10 IDEX 0002), and by SFRI STRAT'US Projects No. ANR-20-SFRI-0012 and No. ANR-17-EURE-0024 under the framework of the French Investments for the Future Program, as well as Grant No. ERC-COG-863487 ``UNIQUE''.

\footnotetext[1]{See Supplemental Material for details about the derivation of the NEGF formalism and the derivation of the current formula, as well as resonant approximation for the polaritonic propagator and derivation of the analytical formula.}

\footnotetext[2]{While the partially self-consistent scheme employed here guarantees current conservation, it does not account for multi-photon emission processes. Such processes are expected to contribute only at higher order, with corrections to the self-energy scaling at least as $(g^2/\kappa)^2$, and should therefore remain negligible for realistic parameter values.}

\bibliography{biblio.bib}

\end{document}

% Use the \preprint command to place your local institutional report
% number in the upper righthand corner of the title page in preprint mode.
% Multiple \preprint commands are allowed.
% Use the 'preprintnumbers' class option to override journal defaults
% to display numbers if necessary
%\preprint{}

%Title of paper
\title{Supplemental Material:Polariton-assisted Inelastic Tunneling through a Quantum Well in the Strong Light-Matter Coupling Regime}

% repeat the \author .. \affiliation  etc. as needed
% \email, \thanks, \homepage, \altaffiliation all apply to the current
% author. Explanatory text should go in the []'s, actual e-mail
% address or url should go in the {}'s for \email and \homepage.
% Please use the appropriate macro foreach each type of information

% \affiliation command applies to all authors since the last
% \affiliation command. The \affiliation command should follow the
% other information
% \affiliation can be followed by \email, \homepage, \thanks as well.
\author{Théophile Seck}
%\email[]{Your e-mail address}
%\homepage[]{Your web page}
%\thanks{}
%\altaffiliation{}
\affiliation{University of Strasbourg and CNRS, CESQ and ISIS (UMR 7006), 67000 Strasbourg, France}
\affiliation{Université de Strasbourg, CNRS, Institut de Physique et Chimie des Matériaux de Strasbourg, UMR 7504, F-67000 Strasbourg, France}
\author{Yanko Todorov}
\affiliation{Laboratoire de Physique et d'Etude des Matériaux, LPEM, UMR 8213, ESPCI Paris, Université, PSL, CNRS, Sorbonne Université, F-75005 Paris, France}
\author{Guido Pupillo}
\affiliation{University of Strasbourg and CNRS, CESQ and ISIS (UMR 7006), 67000 Strasbourg, France}
\author{David Hagenmüller}
\affiliation{Université de Strasbourg, CNRS, Institut de Physique et Chimie des Matériaux de Strasbourg, UMR 7504, F-67000 Strasbourg, France}
%Collaboration name if desired (requires use of superscriptaddress
%option in \documentclass). \noaffiliation is required (may also be
%used with the \author command).
%\collaboration can be followed by \email, \homepage, \thanks as well.
%\collaboration{}
%\noaffiliation

\date{\today}

\begin{abstract}
In this Supplemental Material, we first derive the current formula for a generic superlattice injector and 3DEG extractor and then take the quasi-2D approximation for the miniband injector. Then we detail how the electronic and polaritonic propagators are obtained through an equation of motion method making use of Schwinger functional derivative approach. Using these propagators we expose how, in the polariton-pole approximation for the polaritonic propagator, we can obtain analytical formula for the radiative self-energy, the broadening function and the tunnelling current. Exact results for the radiative self-energy under narrow bandwidth illumination are then presented. Eventually, we make some comments on the pseudoequilibrium distribution, seen as a minimally self-consistent scheme.

\end{abstract}
%\keywords{Suggested keywords}%Use showkeys class option if keyword
                              %display desired
\maketitle

\tableofcontents

\section{Model Hamiltonian \label{Sec:I}}
We consider a two subbands GaAs quantum well (QW) embedded in a wire cavity formed by two gold layers (see Fig.2~(a) of the main text) akin to the one presented in Ref.~\cite{Feuillet-Palma:12}. The cavity mode energy is taken to be resonant with the unique intersubband transition $1 \longrightarrow 2$. The system is contacted by an injector (left contact) described as a superlattice (SL) and an extractor (right contact) described as a 3D electron gas. The SL is modelled as a tight-binding chain of length $L = N\times d$ if $N$ is the number of wells in the SL and $d$ the lattice parameter, in the growth direction \cite{Wacker_2002} and as a free 2D electron gas (2DEG) in plane. 

The QW part of the Hamiltonian reads
\begin{align}
    H_\mathrm{QW} = \sum_{\vect{k},n} \epsilon_n(\vect{k}) c^\dagger_{n \vect{k}} c^{\phantom{\dagger}}_{n \vect{k}}~, \label{eq:HQW}
\end{align}
in which we neglect Coulomb interactions. The fermionic operators $c^{\phantom{\dagger}}_{n \vect{k}} (c^\dagger_{n \vect{k}})$ destroys (creates) an electron in the subband n of the QW with in-plane momentum $\vect{k}$ and energy $\epsilon_n(\vect{k}) = \frac{\hbar^2 k^2}{2m^*} + E_n$ where $E_n = \frac{\hbar^2 n^2\pi^2}{2m^*L_\mathrm{w}^2}$ and $k = |\vect{k}|$. We consider negligible the effect of the applied bias on the QW levels.
The lead are considered perfect leads without interactions, coupled to the QW,
\begin{align}
    &H_\mathrm{leads} = \sum_{\vect{\kp}, \nu}\sum_{k_z = -\pi/d}^{\pi/d} \epsilon_\mathrm{L}(\vect{\kp}, k_z, \nu)d^\dagger_\mathrm{L}(\vect{\kp}, k_z, \nu)d_\mathrm{L}(\vect{\kp}, k_z, \nu)
  +  \sum_{\vect{k}, k_z}\epsilon_\mathrm{R}(\vect{k}, k_z ; V)d^\dagger_\mathrm{R}(\vect{k}, k_z)d_\mathrm{R}^{\phantom{\dagger}}(\vect{k}, k_z) \label{eq:Hsd} \\
    & H_\mathrm{tun} = \sum_{\vect{k},n}\sum_{k_z, \nu} \lambda [c^\dagger_{n\vect{\kp}} d_\mathrm{L}(\vect{k}, k_z, \nu) + d^\dagger_\mathrm{L}(\vect{k}, k_z, \nu)c^{\phantom{\dagger}}_{n\vect{k}}] + \sum_{\vect{k},n} \lambda[c^\dagger_{n\vect{k}} d_\mathrm{R}^{\phantom{\dagger}}(\vect{k}, k_z) + d^\dagger_\mathrm{R}(\vect{k}, k_z)c_{n\vect{k}}^{\phantom{\dagger}}] \label{eq:Htunn}
\end{align}
The fermionic operators $d^{\phantom{\dagger}}_\mathrm{L}(\vect{k}, k_z, \nu) (d^\dagger_\mathrm{L}(\vect{k}, k_z, \nu))$ destroys (creates) an electron in the left contact with in-plane momentum $\vect{k}$ and transverse momentum $k_z$ taken within the miniband $\nu$ of the superlattice, with energy $\epsilon_\mathrm{L}(\vect{k},k_z,\nu,,V) = \frac{\hbar^2 \vect{k}^2}{2m^*} + E_\nu(V) - 2 t_\nu \cos(k_z d)$. Here we include the voltage dependency directly in the bottom of the miniband energy, $E_\nu(V) = E_\nu+ \el V/2$. The left contact bandwidth is thus defined as $D_\mathrm{L} = \mu_0 - E_\nu$ and $\mu_\mathrm{L}(eV) = \mu_0 + \frac{eV}{2}$ is the chemical potential of the left contact. The fermionic operators $d^{\phantom{R}}_\mathrm{R}(\vect{k}, k_z) (d^\dagger_\mathrm{R}(\vect{k}, k_z))$ destroys (creates) an electron in the right contact with in-plane momentum $\vect{k}$, transverse momentum $k_z$ and energy $\epsilon_\mathrm{R}(\vect{k}, k_z, V) = \frac{\hbar^2 (\vect{k}^2 + k_z^2)}{2m^*} + E_\mathrm{c}^\mathrm{R}(eV)$ where $\Rbandedge = \mu_\mathrm{R}(V) - D_\mathrm{R}$ is the energy of the bottom of the conduction band of the right contact, $D_\mathrm{R}$ is the depth of the right contact conduction band and $\mu_\mathrm{R}(V) = \mu_0 - \el V/2$ is the chemical potential of the right contact. 
The conservation of momentum during tunnelling is highlighted in Eq.\eqref{eq:Htunn}, where the 2D momentum $\vect{k}$ is the same between lead operators and QW operators. Note that here we already consider that the tunnelling amplitude $\lambda$ is independent of energy and momentum, which corresponds to a Markovian approximation \cite{HaugJauho2008}. 

The light-matter coupling, in the Coulomb gauge and in the minimal coupling scheme, reads
\begin{align}
    & H_\mathrm{cav} = \hbar \omega_\mathrm{cav}\left(a^\dagger a + \frac 1 2 \right)  \label{eq:Hcav} \\
    & H_\mathrm{I1} = -\im\hbar\sum_{\vect{k}}g_{12}\hat A(c^\dagger_{2\vect{k}} c_{1\vect{k}}^{\phantom{\dagger}} - c^\dagger_{1\vect{k}} c_{2\vect{k}}^{\phantom{\dagger}} ) \label{eq:HI1} \\
    & H_\mathrm{I2} = \frac{\mathcal{N}_\mathrm{el}|\hbar g_{12}|^2}{\ISB} \hat A^2   \label{eq:HI2} \\
    & H_\mathrm{bath} = \sum_p \hbar \omega_{\vect{p}} a^\dagger_{\vect{p}}a_{\vect{p}}^{\phantom{\dagger}} + \sum_{\vect{p}}\mu_{\vect{p}}\hat A_{\vect{p}} \hat A \label{eq:Hbath}\\
\end{align}
where $A = \hat a + \hat a^\dagger$ and $A_{\vect{p}} = \hat a_{\vect{p}}^{\phantom{\dagger}} + \hat a^\dagger_{\vect{p}}$. The Hamiltonian $H_\mathrm{int}$ of the main text Eq.~(1) is defined as $H_\mathrm{int} = H_\mathrm{I1} + H_\mathrm{I2}$ with $D = \frac{\mathcal{N}_{el}|\hbar g_{12}|^2}{\ISB}$.
The bosonic operators $\hat a (\hat a^\dagger)$ destroy (creates) a photon in the cavity mode while $\hat a_{\vect{p}}^{\phantom{\dagger}} (\hat a_{\vect{p}}^\dagger)$ destroy (creates) a photon in the free-space mode $\vect{p}$. The coupling term $H_\mathrm{I1}$ describes the resonant and anti-resonant absorption and emission processes while $H_\mathrm{I2}$ is the diamagnetic term. We add a coupling between the cavity photons and the free-space modes in order to account for losses in the cavity. The full Hamiltonian of the system reads
\begin{equation}
    \mathcal{H}_\mathrm{sys} = H_\mathrm{QW} + H_\mathrm{cav} +H_\mathrm{I1} + H_\mathrm{I2} + H_\mathrm{leads} + H_\mathrm{tun} + H_\mathrm{bath}  \label{eq:Htot}
\end{equation}

The light-matter coupling constant appearing in Eq.~\eqref{eq:HI1} and Eq.~\eqref{eq:HI2} is best expressed using the transition strength and the electrical dipole \cite{Ciuti_Bastard_Carusotto_2005}
\begin{equation}
    \label{eq:coupling_constant_def}
    \hbar g_{12} = \hbar \sqrt{\frac{\el^2f_{12} \sin\theta^2 \omega_{12}}{2 \epsilon_0 m_0 cqV}}
\end{equation}
where we introduced the dimensionless oscillator strength of the transition $f_{12} = \frac{m_0 \omega_{12}d^2_{12}}{\hbar}$ and the electrical dipole of the transition $d_{12} = \bra{1}z\ket{2}$. Here $\theta$ is the incident angle inside the cavity and $V$ is the cavity mode volume, here taken as the volume of the cavity. 

Using Hamiltonian \eqref{eq:Htot} we can now derive an expression for the tunneling current through the system. We follow the usual NEGF approach, as described in \cite{hagenmuller_2017,hagenmuller2018} with light-matter interaction and in the general case in \cite{HaugJauho2008, Pourfath_2014} (for instance). 

\section{NEGF derivation of the tunnelling current \label{section1}}
In the rest of this section we use the notations and conventions $\hbar = 1$, $\partial_a[\cdot] = \frac{\partial [\cdot]}{\partial a}$ the partial derivative w.r.t. variable $a$, and $\delta_f [\cdot] = \frac{\delta[\cdot]}{\delta f}$ the functional derivative w.r.t. function f.

The steady state current flowing through the system can be computed as the expectation value of the time-derivative of the number operator in the left electrode, $J(V) = -\el \langle \partial_t \hat{N}_\mathrm{L} \rangle = -\im \el \langle [H_\mathrm{tot}, \hat N_\mathrm{L}] \rangle $ with $\hat{N}_\mathrm{L} = \sum_{\vect{k}, k_z, \nu} \hat{d}^\dagger_\mathrm{L}(\vect{k}, k_z, \nu)\hat{d}^{\phantom{\dagger}}_\mathrm{L}(\vect{k}, k_z, \nu)$.
The commutator only involves the tunnelling Hamiltonian and calculations lead to the following expression for the current
\begin{equation}
    J(V) = 2 \el \mathfrak{Re}\left[ \sum_{\vect{k}, k_z, \nu} \lambda G^<_\mathrm{L}(\vect{k} , n , t ; \vect{k},k_z,\nu, t')\right]    
\end{equation}
where the mixed QW-Lead Green's function is introduced as $G^<_\mathrm{L}(\vect{k} , n , t ; \vect{k},k_z,\nu, t') = \im \langle \hat d_\mathrm{L}^\dagger(\vect{k},k_z,\nu, t') \hat c_{n\vect{k}}^{\phantom{\dagger}}(t) \rangle$, in Heisenberg picture. To process further we use the equation of motion for the mixed Green's function, by writing $\partial_{\tau'}G_L(\vect{k} , n , \tau ; \vect{k},k_z, \nu, \tau')$, where $G_\mathrm{L}(\vect{k}, n, \tau, ; \vect{k},k_z,\nu, \tau') = {-\im} \langle \mathcal{T}\hat c_{n\vect{k}}^{\phantom{\dagger}}(\tau)\hat d^\dagger_\mathrm{L}(\vect{k},k_z,\nu, \tau') \rangle$ is the contour-ordered Green's function. The imaginary times denoted by the greek $\tau$ are defined on the Keldysh contour. We obtain
\begin{equation}
    \partial_{\tau'} G_\mathrm{L}(\vect{k}, n, \tau ; \vect{k},k_z,\nu, \tau') = -\epsilon_\mathrm{L}(\vect{k}, k_z,\nu ; V) G_L(\vect{k}, n, \tau ; \vect{k},k_z,\nu, \tau') - \sum_{\vect{k}', n'} \lambda G(\vect{k}, n, \tau ; \vect{k}', n', \tau'),
\end{equation}
In the above formula we can already simplify the sum over $\vect{k}'$, using in-plane translational invariance which yields  $G(\vect{k}, n, \tau ; \vect{k}', n', \tau') =  G(\vect{k}, n, \tau ; \vect{k}, n', \tau')\delta_{\vect{k}\vect{k}'}$. This equation of motion can be solved to give
\begin{equation}
    \label{eq:mixte_L_QW_GF}
    G_\mathrm{L}(\vect{k}, n, \tau ; \vect{k},k_z, \nu, \tau') = \int \dif \tau_1 G(\vect{k}, n, \tau ; \eta, \vect{k}, \tau_1)g_\mathrm{L}(\vect{k},k_z,\nu, \tau_1 - \tau')
\end{equation}
Where $g_\eta(\vect{k},k_z, \nu, \tau_1 - \tau') = -\im \langle \mathcal{T} \hat d_\mathrm{L}(\vect{k},k_z,\nu, \tau_1)^{\phantom{\dagger}} \hat d^\dagger_\mathrm{L}(\vect{k},k_z,\nu, \tau') \rangle$ is the bare Green's function of the left lead.
Using Langreth rule for analytic continuation to the real time axis~\cite{HaugJauho2008}, we get the lesser component 
\begin{equation}
    G^<_\mathrm{L}(\vect{k} , n , t ; \vect{k},k_z,\nu, t') = \sum_{n'} \lambda \int dt_1 G^r_{nn'}(\vect{k} , t ; \vect{k} , t_1) g^<_\mathrm{L}(\vect{k},k_z,\nu, t_1 - t') + G^<_{nn'}(\vect{k} , t ; \vect{k} , t_1) g^a_\mathrm{L}(\vect{k},k_z, \nu, t_1 - t'),
\end{equation}
where we introduced the retarded and lesser component of the QW Green's function and the lesser and advanced component of the lead Green's function. Fourier transforming over time and injecting into the current formula we obtain
\begin{equation}
    G^<_\mathrm{L}(\vect{k} , n  ; \vect{k},k_z,\nu, \omega) = \sum_{n'} \lambda \left[ G^r_{nn'}(\vect{k} , \omega) g^<_\mathrm{L}(\vect{k},k_z,\nu, \omega) + G^<_{nn'}(\vect{k} , \omega) g^a_\mathrm{L}(\vect{k},k_z,\nu, \omega) \right],
\end{equation}
and eventually the current reads:
\begin{equation}
    J(V) = 2 \el \mathfrak{Re}\left[ \sum_{\vect{k}, k_z}\sum_{n,n',\nu} \lambda^2  \int \frac{\dif \omega}{2\pi} \left[ G^r_{nn'}(\vect{k} , \omega) g^<_\mathrm{L}(\vect{k},k_z,\nu, \omega) + G^<_{nn'}(\vect{k} , \omega) g^a_\mathrm{L}(\vect{k},k_z,\nu, \omega) \right] \right].
\end{equation}

In principle, the above formula is a sum over all $\{n,n'\}$. As we restrict ourself to the two subband case, the electronic Green's function will be diagonal in the end and we can already simplify the notations. Caution should be exert when doing such simplification, as for $n_\mathrm{\max} = 3$, correlations between the first subband and the third one arise due to radiative self-energy, rendering the Green's function $G_{13}$ non zero. 

Using the fact that the injector electrode is non interacting, we introduce the exact expression of the left contact Green's function $g^<_\mathrm{L}(\vect{k},k_z,\nu, \omega) = 2i\pi\delta\left[\omega - \epsilon_\mathrm{L}(\vect{k},k_z,\nu ; V)\right]f_\mathrm{L}(\omega)$ and $g^a_\mathrm{L}(\vect{k},k_z,\nu, \omega) = \frac{1}{\omega - \epsilon_\mathrm{L}(\vect{k},k_z,\nu;V) - i0^+}$, where $f_{\eta}(\omega; V) = \heaviside{\mu_{\eta}(V) - \omega}$ is the distribution function of the lead $\eta$ at 0K. Here we make explicit the bias dependency of the injector dispersion relation.

We express $g^a_\mathrm{L}(\vect{k},k_z,\nu, \omega) = \frac{1}{\omega - \epsilon_\mathrm{L}(\vect{k},k_z,\nu ; V) - i0^+} =  \im\pi\delta\left[ \omega - \epsilon_\mathrm{L}(\vect{k}, k_z,\nu,V) \right] + p.v. \frac{1}{\omega - \epsilon_\mathrm{L}{(\vect{k},k_z,\nu ; V)}}$ with the Sohotsky formula, and simplifying the real part, we obtain (converting momentum summation into energy integration using the 2D density of states $\DOS = \frac{m^*}{\pi \hbar^2}$ and the tight-binding density of states $\rho_\nu(\Omega) = \frac{L}{\pi d} \frac{1}{\sqrt{4t_\nu^2 - (\Omega - E_\nu(V))^2} }$):
\begin{align}
    J(V) = -2e\lambda^2 & \int\frac{d\omega}{2\pi}  \int \DOS  \dif \epsilon_{k} \int_{E_\nu(V) - 2 |t_\nu|}^{E_\nu(V) + 2 |t_\nu|}\rho_\nu(\Omega)d\Omega   \lambda^2\mathfrak{Im}\Biggl\{\sum_{n}\biggl[G^r_{n}(\epsilon_{k}, \omega) 2\pi\delta(\omega - \epsilon_\mathrm{L}(\epsilon(\vect{k}),\Omega))f_\mathrm{L}( \omega, V) + \\
    & G^<_{n}(\epsilon_{k}, \omega)\pi\delta \left[ \omega - \epsilon_\mathrm{L}(\epsilon(\vect{k}),\Omega) \right] \biggr]\Biggr\}.
\end{align}
Where we had to introduce $\Omega = E_\nu(V) - 2 t_\nu \cos(k_z d)$.
It is always possible to express the lesser Green's function using a local distribution function $f_\mathrm{QW}(\omega ; V)$ \cite{Jauho1994}, in a fluctuation-dissipation form \cite{stefanucci_nonequilibrium_2013}, $G^<_{n}(\epsilon_{k}, \omega) = -2\im f_\mathrm{QW}(\omega ; V) \mathfrak{Im}\left[G^r_{n}(\epsilon_{k}, \omega)\right]$. In principle this local distribution function should be calculated via the Keldysh equation and depends on the full system self-energy. Therefore, solving for it is a self-consistent problem. Assuming here it will only depend on frequency and bias (this will be demonstrated later) we further get
\begin{equation}
    \label{eq:current1}
    J(V) = -2e\lambda^2 \DOS \sum_{n, \nu} \int \frac{\dif\omega}{2\pi} \int \dif \epsilon(\vect{k})  \int_{E_\nu(V) - 2 |t_\nu|}^{E_\nu(V) + 2 |t_\nu|}\rho_\nu(\Omega)\dif\Omega 2\pi \delta(\omega - \epsilon(\vect{k}) - \Omega) \Imag{G^r_{n}(\epsilon(\vect{k}),\omega)}[f_\mathrm{L}(\omega, V) - f_\mathrm{QW}(\omega, V)]
\end{equation}

The Eq.~\eqref{eq:current1} describe the current flowing through the left contact as the difference between the current flowing from the injector to the QW and the current flowing from the QW to the injector. This form is not more usefull \textit{a priori} as the knowledge $f_\mathrm{QW}(\omega, V)$ requires the knowledge of the full lesser Green's function and is hence a self-consistent problem.
The next step is to integrate the current over the transverse contribution to the energy $\epsilon(\vect{k})$, the Dirac delta yield $\epsilon(\vect{k}) = \omega - \Omega$, since it's a positive quantity it yields the constraint $\omega \geq \Omega$. We obtain
 \begin{equation}
    \label{eq:current2}
    J(V) = -2e2 \pi\lambda^2 \DOS \sum_{n, \nu} \int \frac{\dif\omega}{2\pi}\int_{E_\nu(V) - 2 |t_\nu|}^{E_\nu(V) + 2 |t_\nu|}\rho_\nu(\Omega)\dif\Omega \Imag{G^r_{n}(\omega - \Omega,\omega)}[f_\mathrm{L}(\omega, V) - f_\mathrm{QW}(\omega, V)]
\end{equation}
Using the above formula Eq.\eqref{eq:current2} one could derive an expression for the current flowing through the miniband injector in a general case. Here we make a first step toward the Q2DED approximation: we restrict our calculation to the first miniband assuming that the Fermi level lies between the first two miniband and that it is not possible, at any bias, to have alignment between higher energy miniband and a QW subband. In other word we assume that injection/extraction from the other miniband is impossible. Thus we re-label $E_{\nu=1} = E_\mathrm{inj}$, $t_\nu = t$, $\rho_\nu(\Omega) = \rho(\Omega)$ and we discard the summation over miniband index. 
Next, we make two very important approximations: first we neglect the energy variation of the 1D tight-binding density of states (Markovian bath) thus $\rho(\Omega) \sim \rho$ and we introduce the injection rate $\Gamma(\Omega) = 2 \pi \lambda^2 \rho(\Omega)$
This point is non-trivial: any approximation made here on the 1D DOS of the tight-binding model should be kept consistent when calculating the tunnelling self-energy in order to ensure current conservation (see Sec.\ref{Sec:retarded_func} below).
 \begin{equation}
    J(V) = -2e \Gamma_\mathrm{L} \DOS \sum_{n, \nu} \int \frac{\dif\omega}{2\pi}\int_{E_\mathrm{inj}(V) - 2 |t_\nu|}^{E_\mathrm{inj}(V) + 2 |t_\nu|}\dif\Omega \Imag{G^r_{n}(\omega - \Omega,\omega)}[f_\mathrm{L}(\omega, V) - f_\mathrm{QW}(\omega, V)]
\end{equation}

The next step is a very important one in our formalism. As in Ref.~\onlinecite{Jauho1994} we introduce the pseudo-equilibrium approximation for the QW distribution function, $f_\mathrm{QW}(\omega) = \frac{\Gamma_\mathrm{L} f_\mathrm{L}(\omega) + \Gamma_{\mathrm{R}} f_R{\omega}}{\Gamma_\mathrm{L} + \Gamma_{\mathrm{R}}} $, where $\Gamma_\mathrm{R}$ and $\Gamma_\mathrm{L}$ are the tunnelling induced broadening arising from the tunneling self-energies. This can be seen as a minimally self-consistent approximation, as shown below in Sec.~\ref{sec:pseudo_min_cons}. This approximation allows us to take into account the non-equilibrium nature of the current without resorting to a full self-consistent scheme in the current formula.
It is easy to show that within this approximation the current is rewritten 
 \begin{equation}
    \label{eq:Current_f_QW}
    J(V) = -2e \DOS \sum_{n, \nu} \int_\Omega \frac{\dif\omega}{2\pi}\int_{E_\mathrm{inj}(V) - 2 |t_\nu|}^{E_\mathrm{inj}(V) + 2 |t_\nu|}\dif\Omega \frac{\Gamma_\mathrm{L}\Gamma_\mathrm{R}}{\Gamma_\mathrm{L}+\Gamma_\mathrm{R}}\Imag{G^r_{n}(\omega - \Omega,\omega)}[f_\mathrm{L}(\omega, V) - f_\mathrm{R}(\omega, V)]
\end{equation}
A very safe assumption to make is the proportionate coupling to the leads \cite{HaugJauho2008}, assuming $\Gamma_\mathrm{R} = \lambda \Gamma_\mathrm{L}$ we get
 \begin{equation}
    J(V) = -2e \Gamma_{\mathrm{L}} \DOS \sum_{n, \nu} \int_\Omega \frac{\dif\omega}{2\pi}\int_{E_\mathrm{inj}(V) - 2 |t_\nu|}^{E_\mathrm{inj}(V) + 2 |t_\nu|}\dif\Omega \frac{\lambda}{\lambda + 1}\Imag{G^r_{n}(\omega - \Omega,\omega)}[f_\mathrm{L}(\omega, V) - f_\mathrm{R}(\omega, V)]
\end{equation}

We define the asymetry factor $a(\lambda) = \frac{1}{1+1/\lambda}$, in practice we took $\lambda = 1$ in the main text. The form of the current above is not yet the one used in the main text, in order to recover it we need to make explicit the Q2DEG approximation. To highlight it we perform a new change of variable $\tilde \Omega = \Omega - E_\mathrm{inj}(V)$, the differential element is unchanged so $\dif \tilde \Omega = \dif \Omega$ but the limits changes to $$\Omega = E_\mathrm{inj}(V) + 2|t_\nu| \Rightarrow \tilde \Omega = 2|t_\nu|$$ $$\Omega = E_\mathrm{inj}(V) - 2|t| \Rightarrow \tilde \Omega = -2|t|$$ and the current becomes :
\begin{align}
    J(V) = -2 \el \DOS  \Gamma_\mathrm{L} a(\lambda) \sum_n \int_{\Omega + E_\mathrm{inj}(V)} \frac{\dif \omega}{2\pi} \int_{-2|t|}^{2|t|} \dif \tilde \Omega \Imag{G^\mathrm{r}_n(\tilde \Omega + E_\mathrm{inj}(V))}\left[f_\mathrm{L}(\omega) - f_\mathrm{R}(\omega)\right],
\end{align}
Let us now explicit the spectral function. We define $k_0 = \sqrt{\frac{2m^*(\omega - \Omega - E_\mathrm{inj}(V))}{\hbar^2}}$, it reads 
\begin{align}
   -1\times\Imag{G^\mathrm{r}_n(\tilde \Omega + E_\mathrm{inj}(V))} = \frac{\Gamma^\mathrm{tot}_n(k_0, \omega)/2}{(\tilde \Omega + E_\mathrm{inj}(V) - E_n - \Lambda^\mathrm{tot}_n(k_0, \omega))^2 + (\Gamma^\mathrm{tot}_n(k_0, \omega)/2)^2}
\end{align}
where $2\times\Gamma^\mathrm{tot}_n(k_0, \omega) = \Gamma_\mathrm{L} + \Gamma_\mathrm{R} + \Gamma_{\mathrm{int}}(k_0, \omega) = \Gamma + \Gamma_{\mathrm{int}}(k_0, \omega)$. Where $\Gamma_{\mathrm{int}}$ is a general interaction induced broadening.
Our goal is to define a figure-of-merit allowing us to justify the Q2DEG limit. Let us assume that the miniband bandwidth $4|t| = \Delta_\mathrm{SL}$ is very narrow. In the non-interacting case and around a pole (E$_\mathrm{sl}(V) = E_n$) the spectral function reduces to, 
\begin{align}
    -\Imag{G^\mathrm{r}_n}|_\mathrm{pole} \sim \frac{\Gamma/2}{\tilde \Omega^2 + (\Gamma/2)^2} = \frac{1}{\Gamma/2}  \frac{1}{1 + \frac{\tilde \Omega}{\Gamma/2}^2} \simeq \frac{1}{\Gamma/2}  \frac{1}{1 + \frac{\Delta_\mathrm{SL}^2}{\Gamma^2}} \xrightarrow{\makebox[2cm]{$\Delta_\mathrm{SL} \ll \Gamma$}} \frac{1}{\Gamma/2}
\end{align}
Thus if the bandwidth of the miniband is narrower than the total broadening function, we can safely neglect the dependency of $\tilde \Omega$ of the spectral function. In the interacting case, with light-matter interactions, this f.o.m. is modified, in the slow injection limit $\Gamma \ll 2g_{12}^2/\kappa$, $\frac{\Delta_\mathrm{SL}}{2 g_{12}^2/\kappa} \ll 1$

In this limit we neglect $\tilde \Omega$ and $k_0 \rightarrow \sqrt{\frac{2m^*(\omega - E_\mathrm{inj}(V))}{\hbar^2}}$ which imposes the condition $\omega \geq E_\mathrm{inj}(V)$. The integral over $\tilde \Omega$ integrates to $\Delta_{\mathrm{SL}}$ and we end up with :
\begin{align}
    \label{eq:current_final}
    J(V) =  \frac{J_0 a(\lambda)}{2\pi} \sum_n \int_{E_\mathrm{inj}(V)} \frac{\dif \omega}{2\pi} \Imag{G^\mathrm{r}_n(k_0, \omega)}\left[f_\mathrm{L}(\omega) - f_\mathrm{R}(\omega)\right],
\end{align}
Where we defined $J_0 = \frac{-2 \el}{\hbar} \Gamma_\mathrm{L} \DOS \Delta_\mathrm{SL}$ our units of current, weighted by a dimensionless factor related to the miniband bandwitdh and the surface area of the system (via $\DOS$). To recover Eq.~(2) of the main text, we first set $\lambda = 1 \Rightarrow a(\lambda) = \frac 1 2$ and we reintroduce the integral over $\epsilon(\vect{k})$ and $\Omega$, to get : 
\begin{align}
    \label{eq:current_main_text}
    J(V) = \frac{J_\mathrm{0}}{4\pi} \iint F_{\vect{k}}(\omega , V) T_\mathrm{QW}(k, \omega)\left[f_\mathrm{L}(\omega, V) -  f_\mathrm{R}(\omega, V)\right] \mathrm{d}k \mathrm{d} \omega.
\end{align}
where we introduced the supply function of the superlattice, encoding the geometry of the injector, as $F_{\vect{k}}(\omega, V) = \int_{E_\mathrm{inj}(V) - 2|t|}^{E_\mathrm{inj}(V) + 2|t|}\delta\!\left[\omega - \epsilon_L(k, \Omega, V) \right]\dif \Omega$, and the transmission function encoding the tunnelling channels in the well as $T_\mathrm{QW}(k, \omega) = \Imag{G^\mathrm{r}_n(\epsilon(\vect{k}), \omega)}$. The form Eq.~\eqref{eq:current_main_text} is very general, and the current can always be cast with such a function whatever the geometry of the injector. 

\section{Dyson equation for the electronic propagator \label{section2}}
The solution for the electronic Green's function is obtained through the well known equation of motion (e.o.m.) technique, described in any many-body physics textbook \cite{bruus2004manybody, stefanucci_nonequilibrium_2013}. In order to treat the light-matter interaction, we use the functional derivative technic, firest pioneered by Hedin \cite{Strinati_1988, stefanucci_nonequilibrium_2013} and applied to electron-phonon interaction by Schrieffer in Ref.\onlinecite{Engelsberg_Schrieffer_1963}. The following derivation is based on the latter, already applied to light-matter interactions in \cite{hagenmuller_2017,hagenmuller2018}. We derive the equation of motion first for two undefined injector and extractor, only indexed with in-plane momentum $\vect{k}$ (conserved during tunnelling) and internal variable $\sigma$, left purposely undefined, not conserved during tunnelling, in order to keep the derivation general. The following derivation thus can be applied to any injector and extractor. 

We start by adding a source term in the Hamiltonian, $\hat \Phi(t) = \hat A \varphi_2(t)$, that will help us generate the relevant many-body Green's function uncountered during derivation. We only require that this source field vasnishing at $t\longrightarrow \pm \infty$ and that is is Hermitian for all time.
The electronic Green's function is then expressed, in interaction picture \textit{w.r.t the source field}, as $G_{n,n'}(\vect{k},t;\vect{k}',t') = \frac{-\im \brags \mathcal{T}\left\{ \hat S c_{\vect{k},n}(t)^{\phantom{\dagger}}c^\dagger_{\vect{k}',n'}(t') \right\} \ketgs}{\brags \hat S \ketgs}$, with the $\hat S$ matrix defined by $\hat S = \e{\int_{-\infty}^{+\infty}\hat \Phi(\tau)d\tau}$. Note that the Green's function is thus here evaluated on the \textit{total} ground state of the Hamiltonian $\mathcal{H}_\mathrm{tot}$, and the Green's function only depends on the probing field through the $\hat S$ matrix. At the end of the derivation we will implicitly take the limit $\varphi \longrightarrow 0$ where the above Green's function coincide with the physical Green's function of the system.
The e.o.m. for the electronic Green's function is obtained by writing the time-derivative of the time-ordered Green's function $\partial_t G(\vect{k}, n, t ; \vect{k}, n',t')$. Note that rigorously this derivation is made on the Keldysh contour, and thus the time indices here run on the complex time contour. It yields, after calculation of the relevant commutator,
\begin{align}
    \label{eom_QW_final}
    i \partial_t G(\vect{k}, n, t ; \vect{k}, n',t') = &\delta(t - t')\delta_{nn'} + \epsilon_n(\vect{k})G(\vect{k},n,t;\vect{k},n',t') \nonumber \\
    & +  \sum_{\eta, \sigma} \lambda G(\eta, \vect{k},\sigma, t ; \vect{k}, n',t') \nonumber \\
    & -\im \hbar \sum_{n_1} g_{n_1 n} F^{(2)}(t; \vect{k}, n_1,t ; \vect{k}, n',t') \nonumber \\
\end{align}
where we introduced the second order Green's function $F^{(2)}(t_1;\vect{k}, n', t'; \vect{k},n,t) = -\im _0{\Bra{}}\timeO{\hat A(t_1)c^{\phantom{\dagger}}_{n' \vect{k}}(t')c^\dagger_{n \vect{k}}(t)} {\ket{}_0}$
using functional derivative identity, it is possible to show 
\begin{align}
\im \frac{\delta G(\vect{k}, n_1,t ; \vect{k}', n', t')}{\delta \varphi_2(-q_c,t)} &= F^{(2)}(t;\vect{k}, n_1, t ; \vect{k}, n', t') - G(\vect{k}, n_1, t ; \vect{k}, n', t')\braket{\hat A(t)}, \label{eq:func_diff_phi2}
\end{align}
Calculating explicitly the functional derivative in the l.h.s of Eq.~\eqref{eq:func_diff_phi2} using the definition of the elctronic Green's function, an explicit expression for the second order Green's function $F^{(2)}$ is derived, 
\begin{align}
    \label{eq:second_order_GF_2}
    F^{(2)}(t; \vect{k} , n_1,t ; \vect{k}, n',t') = &+\sum_{n_4,\vect{k_4}}g_{n_4,n}\int dt_4 \delta(t - t_4)G(\vect{k_4}, n_4, t_4 ; \vect{k}, n', t')\braket{\hat A{(t)}} \nonumber \\ 
    &+i \sum_{n_1} \sum_{\substack{\vect{k_3}, k4 \\ n_3, n_4}} \iiint dt_3 dt_4 dt_5 g_{n_1 n}g_{n_4,n_3}G(\vect{k}, n_1, t ; \vect{k_3}, n_3, t_3)  \nonumber \\
    & \times \Gamma^{(2)}(t_5; \vect{k_3}, n_3, t_3 ; \vect{k_4}, n_4, t_4)G(\vect{k_4}, n_4, t_4 ; \vect{k}, n', t') D(t_5 ; t).
\end{align}
While the above expression Eq.\eqref{eq:second_order_GF_2} seems very intricated, the interested reader should refer to Ref.~\onlinecite{Engelsberg_Schrieffer_1963,Strinati_1988, stefanucci_nonequilibrium_2013} or any reference on the Schwinger functional derivative technique, to see how these lengthy expressions arise from the functional derivative identities and Green's function identities. 

The first contribution in the r.h.s of Eq.~\eqref{eq:second_order_GF_2} is a mean-field contribution which will enter the non-interacting Green's function (but will eventually vanish) and the second term will define the radiative self-energy. 
We also introduced the vertex function \begin{align}
    \label{eq:vertex_2}
    \Gamma^{(2)}(\vect{k_3}, n_3, t_3 ; \vect{k_4}, n_4, t_4) &= \frac{-1}{g_{n_4,n_3}}\frac{\delta G^{-1}(\vect{k_3}, n_3, t_3 ; \vect{k_4}, n_4, t_4)}{\braket{\delta \hat A(t_5)}},
\end{align}
and the definition of the photonic propagator, $D(t,t') = \frac{\delta \brags \hat A(t)\ketgs}{\delta \varphi(t')}$ \cite{Engelsberg_Schrieffer_1963}.

With these new definitions, the e.o.m. for the electronic Green's function can be rewritten in the Dyson form,

\begin{align}
        \sum_{\vect{k_4},n_4}\int dt_4 \left[ G_0^{-1}(\vect{k},n,t ; \vect{k_4},n_4,t_4)  - \Sigma(\vect{k},n,t ; \vect{k_4},n_4,t_4) \right] G(\vect{k_4},n_4,t_4 ; \vect{k},n',t') = \delta(t-t')\delta_{n,n'}
\end{align}
where the non-interacting Green's function is defined as
\begin{align}
        \label{G0_time}
        G_0^{-1}(\vect{k},n,t ; \vect{k_4},n_4,t_4) = \Bigg[ \left( \im \partial_t - E(\vect{k},n) \right)\delta_{\vect{k},\vect{k_4}}\delta_{n,n_4} +\im g_{n_4,n}\braket{\hat A(t_4)} \Bigg]\delta(t-t_4)
    \end{align}
The ground-state of the system is the vacuum EM field and the term proportional to $\braket{\hat A(t_4)}$ vanishes. The self-energy is given by the sum of the tunnelling self-energy and the radiative self-energy :
\begin{align}
    \Sigma(\vect{k},n,t ; \vect{k_4},n_4,t_4) = \Sigma_\mathrm{tun}(\vect{k},t : \vect{k_4}, t_4) + \Sigma_\mathrm{rad}(\vect{k},n,t ; \vect{k_4},n_4,t_4)
\end{align}

\begin{equation}
    \Sigma_\mathrm{tun}(k,t : \vect{k_4}, t_4) =\sum_{\sigma, \eta} \lambda^2g_\eta(\vect{k},\sigma, t-t_4)\delta_{\vect{k_4},k}
\end{equation}

\begin{equation}
    \begin{split}
        \Sigma_{\mathrm{rad}}(\vect{k},n,t ; \vect{k_4},n_4,t_4) = + \sum_{n_1,n_3}\sum_{\vect{k_3}} \iint dt_3 dt_5 g_{n_1 n}g_{n_4,n_3}G(\vect{k}, n_1, t ; \vect{k_3}, n_3, t_3) \\ \times \Gamma^{(2)}(t_5;\vect{k_3}, n_3, t_3 ; \vect{k_4}, n_4, t_4)D(t_5 ; t).
    \end{split}
\end{equation}

To proceed further, we introduce the Dyson equation in the vertex function to derive,
\begin{align}
    \label{eq:Gamma2_final}
    \Gamma^{(2)}(t_5, \vect{k_3}, n_3, t_3 ; \vect{k_4}, n_4, t_4) = \frac{1}{g_{n_4,n_3}}\left[ -\im g_{n_4,n_3}\delta(t_4-t_5)\delta(t_3-t_4)\delta_{\vect{k_3}-\vect{k_4}} + \frac{\delta \Sigma(\vect{k_3}, n_3, t_3 ; \vect{k_4}, n_4, t_4)}{\delta\braket{\hat A(t_5)}}\right].
\end{align}
To first order (as done in the SCBA in Ref.\onlinecite{hagenmuller_2017}), we discard vertex corrections and only keep the first term in the r.h.s of Eq.~\eqref{eq:Gamma2_final}. We can make use of this expression to calculate the radiative self-energy in frequency space as
\begin{align}
\Sigma_\mathrm{rad}(\vect{k_4},\omega, n,n_4) &= -\im \sum_{n_1,n_3}\int \frac{\dif \omega'}{2\pi} g_{n_1, n}g_{n_4, n_3}G_{n_1,n_3}(\vect{k_4}, \omega')D(\omega'-\omega) \label{eq:Sigma_final_omega}.
\end{align}
In order to get a more compact expression we can first make use of the in-plane translational invariance which reads : $G(\vect{k_4},n_4,t_4 ; \vect{k},n',t') = G(\vect{k},n_4,t_4 ; \vect{k},n',t')\delta_{\vect{k_4},\vect{k}}$. Further simplification can be done by recalling that we work in the 2 subband subspace, and that the coupling constant $g_{n,n'}$ is non zero only for $n\neq n'$. The radiative self-energy then reads (with some convenient change of notation/variables):
\begin{align}
\Sigma_{\mathrm{rad}, n}(\vect{k},\omega) &= +\im g_{12}^2\int  \frac{\dif \omega'}{2\pi} G_{n'}(\vect{k}, \omega' + \omega)D(\omega') \label{eq:Sigma_rad_final},
\end{align}
where $n' = 2$ if $n = 1$ (and reciprocally). The tunnelling self-energy, in frequency space, reads
\begin{equation}
    \label{eq:Sigma_tunn}
    \Sigma_\mathrm{tun}(\vect{k}, \omega ; V) = \lambda^2 \sum_{\sigma, \eta} g_\mathrm{\eta}(\vect{k},\sigma, \omega ; V),
\end{equation}
were we added the explicit dependence in the bias voltage for completion. For a 3D leads, $\sigma = k_z$ and for the SL lead $\sigma = \{\nu, k_z\}$

The Dyson equation is then rewritten in frequency domain as:
\begin{align}
        \sum_{n_4} \left[ G_0^{-1}(\vect{k},\omega, n, n_4)  - \Sigma(\vect{k},\omega, n,n_4) \right] G(\vect{k},\omega, n_4, n') = \delta_{nn'}
\end{align}
In the general case the above equation can be cast in a matrix form by rewritten the quantity as matrix in the subbands indices. In the 2 subbands limit it can be shown that it reduces to the following algebraic equation:
\begin{align}
    \label{eq:Dyson_qw}
    G_n(\vect{k},\omega) = \frac{1}{\left[G_n^0(\vect{k},\omega)\right]^{-1} - \Sigma_n(\vect{k},\omega)} 
\end{align}
beware that the equation Eq.~\eqref{eq:Dyson_qw} is a self-consistent equation as the knowledge of the self-energy requires the knowledge of the Green's function itself, through Eq.\eqref{eq:Sigma_final_omega}. In order to calculate the radiative self-energy, we need to have an expression for the cavity Green's function, which is the subject of the next section.

\section{Dyson equation for the polaritonic propagator \label{section3}}
In order to obtain the cavity Green's function, we start from the definition introduced in Eq.\eqref{eq:vertex_2},
\begin{align}
    \label{eq:def_cavity_greens}
    D(t,t') = \frac{\delta \brags{}\hat A(t)\ketgs{}}{\delta \varphi(t')} = \delta_{\varphi} \braket{\hat A(t)}.
\end{align}
By taking the second derivative of $\hat A(t)$ with respect to time and then taking the functional derivative of the particular expectation value with respect to the vector potential probing field, we get the following e.o.m. for the cavity Green's function
\begin{align}
    \label{eom_cavity_main}
    \left(\frac{1}{2\omega_\mathrm{c}} \partial_{t^2} + \frac{\omega_\mathrm{c}}{2} + 2\frac{N_\mathrm{el}|g_{12}|^2}{\ISB}\right) D(t,t') = -\delta(t-t') + \sum_{\substack{\vect{k} \\ n,n'}} g_{n,n'} \frac{\delta G(n,\vect{k},t ; n',\vect{k},t^+)}{\delta \varphi(t')} \nonumber \\ -\sum_{\vect{p}}\int \dif t" \mu_{\vect{p}}^2D(t ; t") D^0_\mathrm{bath}(\vect{p},t",t').
\end{align}
where we introduced the free-space propagator, $D^0_\mathrm{bath}(\vect{p},t",t') = \brags \timeO{\hat A_{\vect{p}}(t") \hat A_{-\vect{p}}(t')} \ketgs$.

Using the functionnal derivative identity: 
\begin{align}
    \frac{\delta G(1, 2)}{\delta \varphi(5)} = - \iint \mathrm{d}3 \mathrm{d}4 G(1,3) \frac{\delta G^{-1}(3,4)}{\delta \varphi(5)} G(4,2),
\end{align}
where $\int \mathrm{d}1 = \sum_{\vect{k_1}, n_1}\int dt_1$, and a chain rule for the subsequent functional derivative, considering $G^{-1}$ now as a function of both the probing field and the expectation value of the cavity field, we can rewrite the functional derivative in the r.h.s of eq~\eqref{eom_cavity_main} as a function of the vertex function $\Gamma^{(2)}$ and the cavity Green's function $D$
\begin{equation}
    \begin{split}
        \label{eq:func_derivative_cavity}
        \frac{\delta G^{-1}(\vect{k_3},n_3,t_3 ; \vect{k_4},n_4,t_4)}{\delta \varphi_2(\vect{q},t')} = - \sum_{\substack{k \\ \vect{k_3}, \vect{k_4}}}\sum_{\substack{n,n' \\ n_3, n_4}} g_{n,n'}g_{n_4,n_3}G(\vect{k},n,t ; \vect{k_3},n_3,t_5)\Gamma^{(2)}(t_5,\vect{k_3}, n_3, t_3 ; \vect{k_4}, n_4, t_4) \\ \times G(\vect{k_4},n_4,t_4 ; \vect{k}-\vect{q}, n',t^+)D(t_5 ; t').
    \end{split} 
\end{equation}

In order to obtain an e.o.m. in the Dyson form, we need to express the cavity Green's function in Eq.~\eqref{eom_cavity_main} as:
$ D(t,t') = \int \dif t_5 D(t_5, t')\delta(t_5 - t)$, where the order of the variables is chosen to match the photonic propagator in Eq.~\eqref{eq:func_derivative_cavity}. The Dyson equation for the cavity Green's function then reads
\begin{align}
    \label{Dyson:cavity}
    \int \dif t_5 \left[ D_0^{-1}(t_5 ; t') - \Sigma_{\Pi}(t,t_5) - \Sigma_{A^2}(t_5,t) - \Sigma_{bath}(t_5,t') \right] D( t_5 ; t') = + \delta(t-t')
\end{align}
where we defined the free cavity Green's function as:
\begin{align}
    \label{def:D0}
    D_0^{-1}(t_5 ; t') &= \left( \frac{-1}{2\omega_\mathrm{c}} \partial_{t_5^2} - \frac{\omega_\mathrm{c}}{2}\right) \delta(t_5 - t'), 
\end{align}
and the $A^2$ self-energy and cavity-bath self-energy as:
\begin{align}
    \Sigma_{A^2}(t_5, t) &=  + 2 \frac{N_\mathrm{el}|g_{12}|^2}{\Delta_{12}}\delta(t_5 - t'), \nonumber \\
    \Sigma_\mathrm{bath}(t_5, t') &= + \sum_{\vect{p}}\int \mu_{\vect{p}}^2 D^0_{bath}(\vect{p},t_5,t'),
\end{align}
And the polarization Green's function, obtained by simplifying the expression of the second vertex Eq.~\eqref{eq:Gamma2_final}, 
\begin{align}
    \label{Pi2_before_invariance}
    \Sigma_{\Pi}(t,t_5) &= + \im \sum_{\vect{k}, \vect{k_3}}\sum_{\substack{n,n' \\ n_3, n_4}} g_{n,n'}g_{n_4,n_3}G(\vect{k},n,t ; \vect{k_3},n_3,t_5)G(\vect{k_3},n_4,t_5 ; n', t^+). \nonumber \\     
\end{align}
The label \textit{polarization} will be explicited below.
We can now safely take the limit $\varphi_2 \longrightarrow 0$. In this limit, we can take advantage of the time and in-plane translationnal invariance. Also remind that since we only have two subbands, we can get the coupling strength out of the sum and further simplify Eq.~\eqref{Pi2_before_invariance} as
\begin{align}
    \label{Pi2_t_final}
    \Sigma_{\Pi}(t-t_5) &= - \im g_{12}^2 \sum_{\vect{k}}\sum_{n \neq n'}G_{n}(\vect{k},t-t_5)G_{n'}(\vect{k},t^+).
\end{align}
We can now fourier transform w.r.t. $(t-t_5)$ to get 
\begin{align}
    \Sigma_{\Pi}(\omega) = - \im g_{12}^2 \sum_{k}\sum_{n \neq n'} \int \frac{\dif \omega'}{2\pi} G_{n}(k,\omega')G_{n'}(k,\omega' - \omega)= - \im g_{12}^2 \sum_{n \neq n'}\Pi_{nn'}(\omega).
\end{align}
Where we introduced the polarization function $\Pi_{nn'}(\omega) = \sum_{\vect{k}}\int \frac{\dif \omega'}{2\pi} G_{n}(\vect{k},\omega')G_{n'}(\vect{k},\omega' - \omega)$, of the QW.

Finally, we can also re-express the self-energy arising from the coupling to the bath. In frequency domain, converting the summation over free-space 3D momentum to an integral over energy, we can write the cavity-bath self-energy as
\begin{align}
    \Sigma_\mathrm{bath}(\omega) &= \int d\epsilon_{\vect{p}} \kappa D^0_\mathrm{bath}(\epsilon_{\vect{p}},\omega),
\end{align}
where we introduced the cavity photon decay rate $\kappa(\epsilon_{\vect{p}}) = \rho(\epsilon_{\vect{p}})\mu^2 \sim \kappa$, independent of energy in the Markovian approximation. In frequency space, the two remaining contribution in the Dyson equation are given by $D_0( \omega) = \frac{2\omega_\mathrm{c}}{\omega^2 -\omega^2_\mathrm{c}}$ and $\Sigma_{A^2} = 2\frac{N_{c}|\hbar g_{12}|^2}{\omega_{12}}$.

Thus, the complete Dyson equation for the cavity Green's function in frequency space reads:
\begin{align}
    \label{eq:Dyson_cavity}
    D(\omega) = \left[ D^{-1}_0(\omega) - \Sigma_{\Pi}(\omega) - \Sigma_{A^2} - \Sigma_\mathrm{bath}(\omega)\right]^{-1}.
\end{align}

\section{Retarded function and self-energies \label{Sec:retarded_func}}
The Dyson equation for the retarded cavity and electronic Green's function can be obtained directly using the Dyson's equation Eq.\eqref{eq:Dyson_qw} and Eq.\eqref{eq:Dyson_cavity}
\begin{align}
    G^r_n(k,\omega) &= \frac{1}{\left[G_n^{0,\mathrm{r}}(\vect{k},\omega)\right]^{-1} - \Sigma^r_n(\vect{k},\omega)} \label{eq:QW_retarded_Dyson_eq}\\
    D^\mathrm{r}(\omega) &= \left[ [D^{\mathrm{r},0}(\omega)]^-1 - \Sigma_{\Pi}^r(\omega) - \Sigma_{A^2} - \Sigma_\mathrm{bath}^r(\omega)\right]^{-1} \label{eq:Cav_retarded_Dyson_eq}
\end{align}
where the bare retarded Green's function are given by $G^{0,r}_n(\vect{k},\omega) = \frac{1}{\omega - \epsilon_{\vect{k}} - E_n + \im 0^+}$ and $D^{r,0}(\omega) = \frac{2\omega_\mathrm{c}}{(\omega + \im 0^+)^2 -\omega^2_\mathrm{c}}$. The retarded radiative self-energy is calculated using the Kramers-Kronig relation, $\Sigma^r_\mathrm{rad} = \Lambda + \im \frac \Gamma 2$ where $\Gamma$ is the broadening function $\Gamma = \Sigma_\mathrm{rad}^> - \Sigma_\mathrm{rad}^<$ and $\Lambda$ is the energy shift with $\Lambda = \frac{1}{2}\mathcal{H}\left[\Gamma\right]$, where $\mathcal{H}$ is the Hilbert transform. The retarded and advanced component are calculated using the Langreth on Eq.\eqref{eq:Sigma_rad_final} yielding
\begin{align}
    \label{eq:Sigma_rad_lessgtr}
    \Sigma_{\mathrm{rad}, n}^\lessgtr(k, \omega) &= -\im g_{12}^2 \int \frac{\dif\omega'}{2\pi} G_{n'}^{\lessgtr}(k, \omega + \omega')D^{\gtrless}(\omega').
\end{align}
Let us now give an expression for the tunneling self-energies. Starting at Eq.\eqref{eq:Sigma_tunn}, we have for both 3D right lead and for the SL in the Q2DEG limit $\sigma = k_z$.
The retarded SL tunneling self-energy reads, 
\begin{align}
    \label{eq:Sigmar_L}
    \Sigma^r_L(k,\omega) = \sum_{k_z = -\pi/d}^{\pi/d} \lambda^2 g^{r,0}_L(\vect{k}, k_z, \omega) &= \lambda^2 \int_{E_\mathrm{inj}(V) - 2|t|}^{E_\mathrm{inj}(V) + 2|t|} \rho(\Omega) \frac{1}{\omega - \epsilon_{\vect{k}} - \Omega+ i0^+} \dif \Omega \nonumber \\
    & \simeq -\im \pi \lambda^2 \int_{E_\mathrm{inj}(V) - 2|t|}^{E_\mathrm{inj}(V) + 2|t|} \rho(\Omega) \delta(\omega - \epsilon_{\vect{k}} - \Omega ) \dif \Omega \nonumber \\
    &\simeq -\im \frac{\Gamma_\mathrm{L}}{2} F_{\vect{k}}(\omega)
\end{align}
Where we used the Markovian approximation made before. The condition $\epsilon_{\vect{k}} = \omega - \Omega'$ with $\Omega'$ in the bandwidth of the injector, in the current calculation, will yield in Eq.\eqref{eq:Sigmar_L} $\Sigma^r_L(\vect{k},\omega) = -\im \frac{\Gamma_L}{2}\heaviside{\omega - E_\mathrm{inj}} = -\im \frac{\Gamma_L}{2}$ hence justifying the form we took for $f_\mathrm{QW}$ in Eq.\eqref{eq:Current_f_QW}.
The 3D case is worked out analogously, 
\begin{align}
    \Sigma^\mathrm{r}_{\mathrm{R}}(\vect{k},\omega) = \sum_{k_z} \lambda^2 g^{r,0}_\mathrm{R}(\vect{k}, k_z, \omega) = -\im \frac{\Gamma_\mathrm{R}}{2}
\end{align}
Where $\Gamma_\mathrm{R}(\Omega) = 2\pi \rho_{1D}(\Omega) \sim \Gamma_\mathrm{R}$ is the extraction rate, assuming a Markovian bath.

We no turn to the photonic self-energies. The $A^2$ self-energy does not change, as it is an algebraic quantity. The bath self-energy is considered purely imagiary as the coupling to the bath is perturbative. It reads 
\begin{align}
    \Sigma^\lessgtr_{\mathrm{bath}}(\omega) &= \int \dif \epsilon_{\vect{p}} \kappa D^{0,\lessgtr}_\mathrm{bath}(\epsilon_{\vect{p}}, \omega)
\end{align}
using the definition of the bare retarded photonic propagator \cite{Mahan_2000,hagenmuller2018}, $D^{<,0}(\epsilon_{\vect{p}}, \omega) = -2i\pi \left[(N_{\vect{p}} + 1)\delta(\omega + \epsilon_{\vect{p}}) + N_{\vect{p}} \delta(\omega - \epsilon_{\vect{p}})\right]$ and $D^{>,0}(\epsilon_{\vect{p}}, \omega) = -2i\pi \left[(N_p + 1)\delta(\omega - \epsilon_{\vect{p}}) + N_p \delta(\omega + \epsilon_{\vect{p}})\right]$, in the dark (so $N_{\vect{p}} = 0$) we can write the bath self-energy as
\begin{align}
    \Sigma^\lessgtr_{\mathrm{bath}}(\omega) &= -\im\kappa \Theta(\mp\omega)
\end{align}
and hence the retarded self-energy just give some broadening to the photonic propagator, 
\begin{equation}
    \mathfrak{Im}\left[\Sigma^r_{\mathrm{bath}}\right] = -\frac{\kappa}{2} \mathrm{sign}(\omega)
\end{equation}
In the following we incorporate the factor $\frac 1 2$ in the $\kappa$ definition.

We choose to compute the polarization function at zeroth order in the Green's function, the Langreth rule reads
\begin{align}
    \Pi^{r,0}_{n',n}(\omega) = -i \sum_{\vect{k}} \int \frac{d\omega'}{2\pi}G^{<,0}_{n'}(\vect{k}, \omega')G^{a,0}_{n}(\vect{k}, \omega'-\omega) + G^{r,0}_{n'}(\vect{k}, \omega')G^{<,0}_{n}(\vect{k}, \omega'-\omega),
\end{align}
after some simple algebra, and using the definition of the bare retarded and lesser Green's function, we can write the polarization function as
\begin{equation}
    \Pi^{r, 0}_{n',n}(\omega) = \sum_{\vect{k}} \frac{n_{F}(\vect{k},n) - n_{F}(\vect{k},n')}{\omega - \epsilon_{n'}(\vect{k}) + \epsilon_{\vect{k},n} + \im 0^+}.
    \label{Pi_r}
\end{equation}

Using the explicit formula for the leads self-energy, we can now establish rigorously the minimally self-consistent scheme.

\section{Pseudoequilibrium distribution, minimally self-consistent scheme and current conservation. \label{sec:pseudo_min_cons}}
In order to rigorously derive the pseudo-equilibrium distribution let us express the exact lesser Green's function. The Keldysh equation, within our formalism yields (omitting the moment/energy dependency)
\begin{align}
    G^<_n = G^\mathrm{r}_n \Sigma_n^{<,\mathrm{tot}}G^\mathrm{a}_n.
\end{align}
This equation is algebraic and we can regroup $G^\mathrm{r}_nG^\mathrm{a}_n = \frac{A}{\Gamma^\mathrm{tot}}$ where A is the spectral function equal to $-2 \Imag{G^\mathrm{r}_n}$. The lesser Green's function can always be rewritten as 
\begin{align}
    G^<_n = -2i f_\mathrm{QW}\Imag{G^\mathrm{r}_n}, 
\end{align}
where $f_\mathrm{QW} = \Sigma^{<,\mathrm{tot}}_n/\Gamma^\mathrm{tot}_n$ . The full determination of the QW distribution function requires the knowledge of the lesser Green's function and is thus, as already stated, a self-consistent issue. We suppose here that we can neglect the radiatively induced populations and we make the partially self-consistent assumption $\Sigma^\mathrm{tot} = \Sigma_\mathrm{tun}$. This accounts to saying that only tunnelling event (injection/extraction) fixes the QW occupation of states in steady state. Using the fact that $\Sigma_\mathrm{L/R}(\vect{k}, \omega) = -2\im f_\mathrm{L/R}(\omega) \Imag{\Sigma_\mathrm{L/R}(\vect{k},\omega)}$ (true here because we have perfect non-interacting leads), we can express the QW distribution function as
\begin{align}
    f_\mathrm{QW}(\vect{k},\omega,V) = \frac{\Gamma_\mathrm{L}F_{\vect{k}}(\omega, V) f_\mathrm{L}(\omega,V) + \Gamma_\mathrm{R} f_\mathrm{R}(\omega, V)}{\Gamma_{\mathrm{L}}F_{\vect{k}}(\omega, V) + \Gamma_{\mathrm{R}}}.
\end{align}
In the current integration this simplifies in the form used in Eq.\eqref{eq:Current_f_QW}, but in the radiative self-energy it will introduce selection between the lesser and greater component (as explained below in Sec.\ref{Sec:polariton_pole}.) 
We will now show that using this pseudo-equilibrium distribution allows for a pseudo-conservation of the current. 

The usual figure of merit for current conservation is $J_\mathrm{L} + J_\mathrm{R} = 0$ \cite{HaugJauho2008, Pourfath_2014}, stating that the injected current from the left is equal to the one from the right. 

The right injected current can be derived the same way as the left injected current, and the following formula is obtained
\begin{align}
    J_\mathrm{R}(V) = \frac{-2e \DOS \Gamma_\mathrm{R}}{\hbar} \int  \frac{\dif \omega}{2 \pi} \int_{\Rbandedge} \dif \Omega_z \int \dif \epsilon_{\vect{k}} \delta\left[\omega - \epsilon_{\vect{k}} - \Omega_z  \right] \Imag{G^\mathrm{r}_n(\epsilon(\vect{k}), \omega)}\left[f_\mathrm{R}(\omega, V) - f_\mathrm{QW}( V)\right], 
\end{align}
where we made the change of variable $\Omega_z = \frac{\hbar^2k_z^2}{2m^*}$ and then $\Omega_z = \Omega_z + \Rbandedge$. In the above equation, we do not specifiy the exact dependency of the QW distribution function ; a priori it depends on every possible internal variables $(\omega, \epsilon_{\vect{k}}, n)$. As seen above, the QW distribution function is expressed through the self-energy. As we discard the radiative self-energy in this expression, we can readily discard the dependency on $n$. We then rewrite the current within the pseudo-equilibrium approximation as
\begin{align}
    J_\mathrm{R}(V) = \frac{-2e \DOS}{\hbar} \int \frac{\dif \omega}{2 \pi} \int_{\Rbandedge} \dif \Omega_z \int \dif \epsilon(\vect{k})  \frac{\Gamma_{\mathrm{L}}F_{\vect{k}}(\omega, V)\Gamma_{\mathrm{R}}}{\Gamma_{\mathrm{L}}F_{\vect{k}}(\omega, V) + \Gamma_{\mathrm{R}}}\delta\left[\omega - \epsilon_{\vect{k}} - \Omega_z  \right] \Imag{G^\mathrm{r}_n(\epsilon_{\vect{k}}, \omega)}\left[f_\mathrm{R}(\omega, V) - f_\mathrm{L}(\omega, V)\right].
\end{align}

Let us express the injector supply function, it reads
\begin{align}
    F_{\vect{k}}(\omega, V) = \int_{E_\mathrm{inj}(V) - 2|t|}^{E_\mathrm{inj}(V) + 2|t|} \delta[\omega - \epsilon(\vect{k}) - \Omega]\dif\Omega.
\end{align}
Which is non-zero only if the miniband Dirac pole is integrated completely. Let us see what this quantity becomes inside the current integral when we introduce $\epsilon(\vect{k}) = \omega -  \Omega_z$ :
\begin{align}
    F_{\vect{k}}(\Omega_z, V) = \int_{E_\mathrm{inj}(V) - 2|t|}^{E_\mathrm{inj}(V) + 2|t|} \delta[\Omega_z - \Omega]\dif\Omega,
\end{align}
and now the condition for the Dirac integration to be 1 is $\Omega_z \in [E_\mathrm{inj}(V) - 2|t|, E_\mathrm{inj}(V) + 2|t|]$. Assuming that $E_\mathrm{inj}(V) \geq \Rbandedge$ we can rewrite (performing integration over $\epsilon_{\vect{k}}$), 
\begin{align}
    J_\mathrm{R}(V) = \frac{-2e \DOS}{\hbar} \int  \frac{\dif\omega}{2 \pi} \int_{E_{inj}(V) - 2|t|}^{E_{inj}(V) + 2|t|} \dif \Omega_z  \frac{\Gamma_{\mathrm{L}} \Gamma_{\mathrm{R}}}{\Gamma_{\mathrm{R}} + \Gamma_{\mathrm{L}}}\Imag{G^\mathrm{r}_n(\omega - \Omega_z, \omega)}\left[f_\mathrm{R}(\omega, V) - f_\mathrm{L}( V)\right] = - J(V)
\end{align}
where we recovered exaclty the formula Eq.\eqref{eq:Current_f_QW}. Eventually we see that our partially self-consistent scheme allows for a pseudo-conservation of the current. We call it "pseudo-conservation" because we verify the relation $J_L + J_R = 0$, but the total self-energy employed in the calculation is not $\Phi$-derivable and thus is not conserving in the sense of Baym (this could be easily verify as our self-energy does not verify the usual conservation condition see \textit{e.g} Ref.\onlinecite{HaugJauho2008}.) Nevertheless we argue that our approximation scheme yields only physical contributions to the current.

\section{Polariton-pole approximation \label{Sec:polariton_pole}}
The idea of this derivation is to first write the exact expression for the photonic propagator, and then to express it as a sum of poles, located at the polariton energies.
We begin by re-writing the polarization self-energy. With only two subbands, the only non-zero terms in the retarded polarization reads
\begin{align}
    \Pi^{\mathrm{r},0}_{12}(\omega) = \frac{\sum_{\vect{k}} n_\mathrm{F}(\vect{k},1) - n_\mathrm{F}(\vect{k},2)}{\omega - \ISB + i0^+} \nonumber \\
    \Pi^{\mathrm{r},0}_{21}(\omega) = \frac{\sum_{\vect{k}} n_\mathrm{F}(\vect{k},2) - n_\mathrm{F}(\vect{k},1)}{\omega + \ISB + i0^+}
\end{align}
This allows to write an explicit expression for the photonic self-energy (neglecting the vanishing imaginary part) at resonnance 
\begin{align}
    \Sigma_\Pi^r(\omega) = \frac{2\Omega_{12}^2\Delta_{21}}{\omega^2 - \Delta_{21}^2}
\end{align}
where introduced the polaritonic gap at resonnance, ie the Rabi splitting, 
\begin{align}
    \label{eq:Rabi_splitting_def}
    \Omega_{nn'} &= g_{12}S \sqrt{n_\mathrm{el}(n) - n_\mathrm{el}(n')}.
\end{align}

The first step is to rewrite the free photon propagator be including the $A^2$ self-energy in the definition of the cavity mode. With only one ISB, the $A^2$ term is rewritten $\Sigma_{A^2} =  + 2D$ with $D = \frac{\mathcal{N}_{el} g_{12}^2(q_0)}{\Delta_{21}}$ and the free photon propagator reads
\begin{align}
    \left[\tilde D^{\mathrm{r},0}\right]^{-1} = \frac{(\omega - \im 0^+)^2 - \tilde \omega_c^2}{2\tilde \omega_c} \times \frac{\tilde \omega_c}{\omega_c},
\end{align}
where we introduced the renormalized cavity frequency $\tilde \omega_c = \sqrt{\omega_c^2 + 4\omega_c D}$.
Then one can write explicitly the photon Green's functions
\begin{align}
    D^\mathrm{r}(\omega) &= \frac{1}{\left[\tilde D^{\mathrm{r},0}(\omega)\right]^{-1} - \Sigma^\mathrm{r}_\Pi(\omega) - \Sigma^r_\mathrm{bath}(\omega)} \nonumber \\
    &= \frac{1}{\frac{\tilde \omega_c}{\omega_c}\frac{\omega^2 - \tilde \omega_c^2}{2 \tilde \omega_c} - \frac{2\Omega_{12}^2\Delta_{21}}{\omega^2 - \Delta_{21}^2} + \im \kappa \mathrm{sgn}(\omega)} \nonumber \\
    &= \frac{2 \Delta_{21}(\omega^2 - \Delta_{21}^2)}{(\omega^2 - \pplustilde^2)(\omega^2 - \pminustilde^2) + 2\im \kappa \mathrm{sgn}(\omega)\Delta_{21}(\omega^2 - \Delta_{21}^2)}
\end{align}
With the polariton dispersion in USC (at resonnance, $\omega_\mathrm{c} = \ISB$), $\psigmatilde = \sqrt{\Delta_{21}^2 +  2\Omega_{12}^2 +\sigma \sqrt{(\Delta_{21}^2 + 2\Omega_{12}^2)^2 - \Delta_{21}^2}}$. We now perform the polariton-pole approximation : the propagator is separated between it's value around the upper polaritonic peak and the lower polaritonic peak (both in the positive and negative part of the spectrum), and we obtain
\begin{align}
    \label{eq:polariton_pole_approx}
    D^\mathrm{r}(\omega) &= \sum_\sigma \frac{\Zsigtilde }{(\omega - \psigma) + \im \Zsigtilde \kappa} - \frac{\Zsigtilde }{(\omega + \psigma) + \im \Zsigtilde \kappa}.
\end{align}
Where we introduced the dimensionless photonic weight $\Zpmtilde = \frac{\ISB\left(1 \pm \frac{D}{\sqrt{D^2 + \Rabi^2}}\right)}{2\ppmtilde}$ and we also separated between the positive and negative part of the spectrum. The propagator Eq.~\eqref{eq:polariton_pole_approx} has now a dependency on the electronic density in the well. If we had a finite electronic lifetime in the well, the polarization function would have inherited a finite imaginary part that would have also been present in the cavity dressed propagator. For this reason, we relabel the dressed cavity propagator as the \textit{polaritonic} propagator, in order to reflect this hybrid character, $D^\mathrm{r}(\omega) \longrightarrow P^\mathrm{r}(\omega)$.

We will now leverage this approximation to obtain analytical formula for the lesser and greater radiative self-energies. Using the expression of the lesser and greater radiative self-energy Eq.~\eqref{eq:Sigma_rad_lessgtr} with the polaritonic propagator we get
\begin{align}
    \label{eq:Sigma_rad_lessgtr_polariton}
    \Sigma_{\mathrm{rad}, n}^\lessgtr(\vect{k}, \omega) &= -\im g_{12}^2 \int \frac{\dif\omega'}{2\pi} G_{n'}^{\lessgtr}(\vect{k}, \omega + \omega')P^\gtrless(\omega')
\end{align}
There is two propagators in this expression. We will use the bare electronic propagator, $G^{0,<}_n(\vect{k},\omega) = +2\im \pi n_\mathrm{F}(\vect{k},n) \delta(\omega - \epsilon(\vect{k},n))$ and $G^{0,>}_n(\vect{k},\omega) = -2\im \pi [1 - n_\mathrm{F}(\vect{k},n)] \delta(\omega - \epsilon_{\vect{k},n})$ and the polaritonic propagator, expressed as : 
\begin{align}
    P^\lessgtr(\omega)& = \mp 2 \im \Theta(\mp\omega) \Imag{P^r(\omega)} \nonumber \\
\end{align}
where the imaginary part of the retarded polaritonic progagator is given by Eq.\eqref{eq:polariton_pole_approx}. Noting that the Heaviside select for only for positive (negative) frequencies in the greater (lesser) function allowing to rewrite, 
\begin{align}
    \label{eq:USC_photon_D}
    P^\lessgtr(\omega) &= +2\im \Theta(\mp \omega)\sum_\sigma \Zsigtilde \frac{\kappa \Zsigtilde}{(\omega \mp \psigmatilde)^2 + (\kappa \Zsigtilde)^2}
\end{align}

Now we can perform the integration over $\omega'$ in Eq.\eqref{eq:Sigma_rad_lessgtr_polariton}. The delta function in $G^<$ yields $\omega' - \omega = \epsilon_{n'}(\vect{k})$
\begin{align}
    \label{eq:2SB_radiative_self_energy_analytic}
    \Sigma_{\mathrm{rad},n}^<(\vect{k},\omega) = +2\im g_{12}^2n_\mathrm{F}(\vect{k},n')\sum_\sigma &\Zsigtilde \frac{\kappa \Zsigtilde  \Theta(\epsilon_{n'}(\vect{k}) - \omega)}{(\epsilon_{n'}(\vect{k}) - \omega - \psigmatilde)^2 + (\kappa \Zsigtilde)^2} \nonumber \\
    \Sigma_{\mathrm{rad},n}^>(\vect{k},\omega) = -2\im  g_{12}^2 \bar n_\mathrm{F}(\vect{k},n')\sum_\sigma &\Zsigtilde \frac{\kappa \Zsigtilde\Theta(\omega - \epsilon_{n'}(\vect{k}))}{(\epsilon_{n'}(\vect{k}) - \omega + \psigmatilde)^2 + (\kappa \Zsigtilde)^2}
\end{align}

\section{Spectral representation and computation of the energy shift \label{spectral_shift}}
The following calculation finds its roots in the Appendix O of Ref.\onlinecite{stefanucci_nonequilibrium_2013}. Any self-energy can be decomposed into its real and imaginary, $\Sigma(\omega) = \Sigma_1(\omega) + \im \Sigma_2(\omega)$, the two being linked by Kramers-Kronig relations. In that case, the self-energy can be rewritten using only the imaginary part, the so called spectral representation, which reads
\begin{equation}
    \label{eq:Kramer_Kronig}
    \Sigma(\omega) = -\frac{1}{\pi}\int_{-\infty}^{+\infty}\dif \omega' \frac{\Sigma_2(\omega')}{\omega - \omega' + \im \eta} = \frac{1}{2\pi}\int_{-\infty}^{+\infty}\dif \omega' \frac{\Gamma(\omega')}{\omega - \omega' + \im \eta}.
\end{equation}
The function appearing in the second integral is the broadening function, defined earlier as $\Gamma = \im \left[ \Sigma^> - \Sigma^< \right] = -2\mathfrak{Im}\left[\Sigma^r \right]$. It is always possible to decompose $\Gamma = \im \left[ \Sigma^> - \Sigma^< \right] = \Gamma^> + \Gamma^<$ if we define $\Gamma^> = \im \Sigma^>$ and $\Gamma^< = -\im \Sigma^<$. The following calculation is done only for the greater component $\Sigma_n^>$, but it is easily generalisable to the lesser one.
The integration of the greater component reads
\begin{align}
    \label{eq:SigmaR_contour_start}
    \frac{1}{2\pi}\int_{-\infty}^{+\infty}\frac{\im\Sigma^>_{\mathrm{rad},n}(\vect{k}, \omega')}{\omega - \omega' + i\eta} = +2 g_{12}^2 \bar n_\mathrm{F}(\vect{k},n')\sum_\sigma \Zsigtilde \int \frac{\dif \omega'}{2\pi} \frac{\kappa \Zsigtilde\Theta(\omega -\epsilon_{n'}(\vect{k}))}{(\epsilon_{n'}(\vect{k}) - \omega + \psigmatilde)^2 + (\kappa \Zsigtilde)^2}\frac{1}{\omega - \omega' + \im \eta}, 
\end{align}
by rewriting the Lorentzian denominator as $ \omega^2 + \kappa^2 = (\omega + i\kappa)(\omega - i\kappa)$, and performing the subsequent contour integration we obtain (in the limit $\eta \rightarrow 0+$)
\begin{align}
    \label{eq:energy_shift_analytic}
        \Lambda^\mathrm{>}_{n}(\vect{k}, \omega) &= g_{12}^2 \sum_\sigma \Zsig \frac{\omega - (\epsilon_{n'}(\vect{k}) + \psigma)}{(\omega - (\epsilon_{n'}(\vect{k}) + \psigma))^2 + (\kappa \Zsig)^2} \\
        \Lambda^\mathrm{<}_{n}(\vect{k}, \omega) &= g_{12}^2 \sum_\sigma \Zsig \frac{\omega - (\epsilon_{n'}(\vect{k}) - \psigma)}{(\omega - (\epsilon_{n'}(\vect{k}) - \psigma))^2 + (\kappa \Zsig)^2},
\end{align}
were we added the greater contribution for completion. The total energy shift is thus given by $\Lambda_n(\vect{k}, \omega) = \Lambda^>_n(\vect{k}, \omega) + \Lambda^<_n(\vect{k}, \omega) = \Real{\Sigma^\mathrm{r}_{\mathrm{rad},n}(\vect{k}, \omega)}$.

Using the analytical broadening function given by Eq.~\eqref{eq:2SB_radiative_self_energy_analytic} and the analytical energy shift given by Eq.~\eqref{eq:energy_shift_analytic}, we can now compute both an expression for the retarded Green's function and the dark current. 

\section{Analytical formula for the current in the dark \label{dark_current}}
The idea of our analytical derivation is to first dress the photonic propagator to obtain the polaritonic propagator, with which we calculated analytical formula for the radiative self-energy. Then we inject this self-energy in the Dyson equation for the electronic Green's function to obtain an analytical expression for the retarded Green's function. Finally, we inject this retarded Green's function in the current formula Eq.~\eqref{eq:current_final} to obtain an analytical expression for the current in the dark.

We start back at Eq.~\eqref{eq:current_final}. then we calculate the Heaviside arising from the left and right chemical potential, and we introduce the lower bound of the integral $\omega_\mathrm{min} = \max{(E_\mathrm{inj}(V), \mu_\mathrm{R}(V))}$
\begin{equation}
    \label{eq:2SB_current_definition}
    J(V) = \frac{J_{0}}{4 \pi} \sum_n \int_{\omega_\mathrm{min}(V)}^{\mu_\mathrm{L}(V)} \dif\omega -\Imag{G^\mathrm{r}_{n}(\omega, V)},
\end{equation}
Next we explicit the imaginary part of the Green's function using the Dyson equation \eqref{eq:QW_retarded_Dyson_eq}, with the definition of the non-interacting retarded QW Green's function, we get
\begin{equation}
    \Imag{G^r_n(k, \omega)} = \frac{\Gamma_n(k, \omega)/2}{\left[\omega - \epsilon_{k,n} - \Lambda_n(k,\omega)\right]^2 + \left[\frac{\Gamma_n(k,\omega)}{2}\right]^2},  
\end{equation}
taking into account the conservation of energy and in-plane momentum during tunnelling we use $\omega - \epsilon_n(k_0(V)) = \omega - (\omega - E_\mathrm{inj}(V)) - E_n = E_\mathrm{inj}(V) - E_n$, to rewrite 
\begin{equation}
    \Imag{G^r_n(V, \omega)} = \frac{\Gamma_n(V, \omega)/2}{\left[E_\mathrm{inj}(V) - E_{n} - \Lambda_n(V,\omega)\right]^2 + \left[\frac{\Gamma_n(V,\omega)}{2}\right]^2},  
\end{equation}
The bias-dependent broadening function and energy shift function are obtained by injecting the expression of $\epsilon_n(k_0(V)) = \omega - E_\mathrm{inj}(V) + E_n$ in Eq.~\eqref{eq:energy_shift_analytic} and Eq.~\eqref{eq:2SB_radiative_self_energy_analytic} and switching the equilibrium distribution function by the pseudoequilibrium ones. The total broadening function can be separated into the different contribution $\Gamma_n(\vect{k}, \omega) = \Gamma_\mathrm{R} + \Gamma_\mathrm{L} + \Gamma_{\mathrm{rad}, n}(\vect{k}, \omega) = 2 \Gamma_\mathrm{L} + \Gamma_{\mathrm{rad}, n}(\vect{k}, \omega) $
and hence we define the elastic and inelastic contributions to the current
\begin{align}
    J(V) &= J_\mathrm{el}(V) + J_\mathrm{in}(V) \nonumber \\
    J_\mathrm{el}(V) &= \frac{J_{0}}{4\pi} \sum_n \int_{\omega_\mathrm{min}(V)}^{\mu_\mathrm{L}(eV)} \dif\omega \frac{\Gamma_\mathrm{L}}{\left[ E_\mathrm{inj}(V) - E_n - \Lambda_n(V,\omega)\right]^2 + \left[\frac{\Gamma_n(V,\omega)}{2}\right]^2}   \nonumber \\
    J_\mathrm{in}(V) &= \frac{J_{0}}{4\pi} \sum_n \int_{\omega_\mathrm{min}(V)}^{\mu_\mathrm{L}(eV)} \dif\omega \frac{\Gamma_{\mathrm{rad}, n}(V, \omega)/2}{\left[ E_\mathrm{inj}(V) - E_n - \Lambda_n(V,\omega)\right]^2 + \left[\frac{\Gamma_n(V,\omega)}{2}\right]^2}.
\end{align}

It is useful now to give the explicit, pseudo-equilibrium and bias-dependent expression of the radiative broadening function

\begin{align}
    \label{eq:Gamma_rad_V_omega}
    \Gamma_{\mathrm{rad},n}(V, \omega) &= +4\pi g_{12}^2 \sum_\sigma \Zsigtilde \frac{\bar f_\mathrm{QW}(V,\omega,n')\kappa \Zsigtilde  \heaviside{E_\mathrm{inj}(V) - E_{n'}}}{(E_\mathrm{inj}(V) - (E_{n'} + \psigmatilde))^2 + (\kappa \Zsigtilde)^2} + \Zsigtilde \frac{f_\mathrm{QW}(V,\omega, n')\kappa \Zsigtilde  \Theta(E_{n'} - E_\mathrm{inj}(V))}{(E_\mathrm{inj}(V) - (E_{n'} - \psigmatilde))^2 + (\kappa \Zsigtilde)^2} \nonumber \\
    &= \Gamma_{\mathrm{rad}, n}^>(V, \omega) + \Gamma_{\mathrm{rad}, n}^<(V, \omega).
\end{align}

In the above expression, we encounter the pseudo-equilibrium distribution of the QW taken at a given state $(k,n')$. It is expressed as
\begin{align}
    f_\mathrm{QW}(k,\epsilon_{n'}(\vect{k}), V) = \frac{\Gamma_\mathrm{L}F_{\vect{k}}(\epsilon_{n'}(\vect{k}),V)f_\mathrm{L}(\epsilon_{n'}(\vect{k}),V) + \Gamma_\mathrm{R}f_\mathrm{R}(\epsilon_{n'}(\vect{k}),V)}{\Gamma_\mathrm{L}F_{\vect{k}}(\epsilon_{n'}(\vect{k}),V) + \Gamma_\mathrm{R}}.
\end{align}
When doing the replacement $k \longrightarrow k_0(V)$ this no longer depend on $\vect{k}$, hence the notation of Eq.~\eqref{eq:Gamma_rad_V_omega}.
To see how this function introduces some selection on the contribution to the broadening function let us first recall that we are always in a positive bias situation, such that within our parameter range it is verified that $f_\mathrm{R}(\epsilon_{n'}(\vect{k})) = 0 $. Let us now examine the supply function $F_{\vect{k}}(\epsilon_{n'}(\vect{k}), V)$
\begin{align}
    F_{\vect{k}}(\epsilon_{n'}(\vect{k}), V) &= \int_{E_\mathrm{inj} - 2 |t|}^{E_\mathrm{inj} + 2 |t|} \delta(\epsilon_{n'}(\vect{k}) - \epsilon(\vect{k}) - \Omega) d\Omega \nonumber \\
    &= \int_{E_\mathrm{inj} - 2 |t|}^{E_\mathrm{inj} + 2 |t|} \delta(E_{n'} - \Omega) d\Omega \nonumber \\
    &= 1\,\mathrm{if} \, (E_{n'} - E_\mathrm{inj}) \leq 4|t| \nonumber \\
    &= 0\, \mathrm{else}.
\end{align}
within the Q2DEG approximation, we see that both condition $E_\mathrm{inj} \simeq E_{n'}$ and $E_\mathrm{inj} \simeq E_{n'} \pm \omega_\sigma$ can never be simultaneously satisfied. Thus $f_\mathrm{QW}$, within our parameter range, is always zero around the poles of the radiative broadening function.
The only dependency of Eq.\eqref{eq:Gamma_rad_V_omega} with the frequency is through the distribution function, $f_\mathrm{QW}(V,\omega,n')$, obtained via replacement of $\epsilon_{n'}(k_0(V)) = \omega - E_\mathrm{inj}(V)$. We need to calculate to contribution frequency space in the form, 
\begin{align}
    f^>_{n'}(V) &= \int^{\mu_\mathrm{L}(V)}_{\omega_\mathrm{min}(V)} \dif \omega \bar f_\mathrm{QW}(V,\omega - E_\mathrm{inj}(V) + E_{n'}) \nonumber \\
    &= \mu_L(V) - \omega_\mathrm{min}(V) \nonumber \\
    &= f^\mathrm{el}(V).
\end{align}

We can now write the explicit formula of the elastic and inelastic current
\begin{align}
    J_\mathrm{in}(V) &= \frac{J_{0}}{4\pi}f^\mathrm{el}(V) \sum_{n,\alpha} \frac{\Gamma^>_{\mathrm{rad}, n}(V)}{\left[ E_\mathrm{inj}(V) - E_n - \Lambda^>_n(V)\right]^2 + \left[\frac{\Gamma_{\mathrm{tot}, n}(V)}{2}\right]^2}, 
\end{align}
\begin{align}
    J_\mathrm{el}(V) &= \frac{J_{0}}{4\pi} f^\mathrm{el}(V)\sum_n \frac{\Gamma_\mathrm{L}}{\left[ E_\mathrm{inj}(V) - E_n - \Lambda_n(V)\right]^2 + \left[\frac{\Gamma_n(V)}{2}\right]^2}, \\
    \mathrm{where}~f^\mathrm{el}(V) &= \mu_\mathrm{L}(V) - \max(\mu_{\mathrm{R}}(V), E_\mathrm{inj}(V)) \nonumber \\
    &= \el V~\mathrm{if}~\mu_{\mathrm{R}}(V) > E_\mathrm{inj}(V) \nonumber \\
    &= D_\mathrm{L}~\mathrm{else}. \nonumber
\end{align}

We make here two comment that are highlighted in the main text. First, in the high bias regime where we are working on the features always appear at a bias such that $eV > D_\mathrm{L}$ and thus $f^\mathrm{el}(V) = f^>_{n'}(V) = D_\mathrm{L}$. Secondly, in the parameter range used in main text, we are in a very weak USC of order $\Omega \sim 0.1\Delta$. In that regime, the factor $$Z^\sigma \sim \frac{\Delta}{2\omega_{\sigma}} = \frac{\Delta}{2(\Delta \pm \Omega)} \sim \frac 1 2 .$$ Inserting these two remark into the above equations allows to recover the equations presented in the main text. 

\section{Radiative self-energy under illumination \label{Sec:illuminated_current}}
We want to calculate the radiative self-energy under a coherent illumination, we suppose that a laser generates a coherent state in the mode labelled $\vect{L}$ of the free-space. The calculation of the lesser and greater photonic propagators evaluated in a coherent state at the laser energy yields the following expressions
\begin{align}
    \label{eq:photon_propagator_coherent_state}
    D^<_{p}(\omega) &= -2\im \pi \left[ |\alpha_\mathrm{L}|^2\delta_{p,L}  \delta(\omega - \omega_\mathbf{p}) + (|\alpha_\mathrm{L}|^2\delta_{p,L}  + 1) \delta(\omega + \omega_\mathbf{p})\right] \nonumber \\
    D^>_{p}(\omega) &= -2\im \pi \left[(|\alpha_\mathrm{L}|^2\delta_{p,L}  + 1) \delta(\omega - \omega_\mathbf{p}) + |\alpha_\mathrm{L}|^2 \delta_{p,L} \delta(\omega + \omega_\mathbf{p})\right]~.
\end{align}
where we recognized the form of the lesser/greater free-space photonic propagator with non zero number of photons, restricted to the mode $p = L$ via the Kronecker delta $\delta_{p,L}$. Eq.~\eqref{eq:photon_propagator_coherent_state} yield the following bath self-energy for the cavity photons 
\begin{align}
    \label{eq:Sigma_coherent_state}
    \Sigma^<_\mathrm{pump}(\omega) &= -\im \kappa \left[ \Theta(\omega)|\alpha_\mathrm{L}|^2  \frac{\delta(\omega - \omega_\mathrm{L})}{\rho_\mathrm{fs}(\omega_\mathrm{L})} + \Theta(-\omega)(|\alpha_\mathrm{L}|^2  + 1) \frac{\delta(\omega + \omega_\mathrm{L})}{\rho_\mathrm{fs}(\omega_\mathrm{L})}\right] \nonumber \\
    \Sigma^>_\mathrm{pump}(\omega) &= -\im \kappa \left[ \Theta(\omega)(|\alpha_\mathrm{L}|^2\frac{\delta(\omega - \omega_\mathrm{L})}{\rho_\mathrm{fs}(\omega_\mathrm{L})}  + 1)  + \Theta(-\omega)|\alpha_\mathrm{L}|^2  \frac{\delta(\omega + \omega_\mathrm{L})}{\rho_\mathrm{fs}(\omega_\mathrm{L})}\right]~,
\end{align}
where we introduced the free-space density of states $\rho_\mathrm{fs}(\omega_\mathrm{L})$ when going from the discrete sum over the free-space modes to an energy integral. 

We factorise the free-space density of states as $\rho_\mathrm{fs}(\omega) = \frac{V \omega^2}{\pi^2c^3} = N_\perp(\omega) \times \rho_{1D}$ with $\rho_{1D} = \frac{L}{2\pi c}$ and $N_\perp(\omega) = \frac{2A^2 \omega^2}{\pi c^2}$. Here the quantification volume $V = A\times L$ is still undefined. We identify A with the laser beam cross-section $w_0^2$ and thus only $L$ remains undefined. We introduce a longitudinal photon flux $\Phi = \frac{|\alpha_\mathrm{L}|^2 c}{L}$ and thus we identify the power of the CW laser source as $P_\mathrm{L} = I_\mathrm{L} \times w_0^2 = \hbar \omega_\mathrm{L} \Phi$ and we rewrite the term $\frac{|\alpha_L|^2}{\rho_\mathrm{fs}(\omega_\mathrm{L})} = \frac{2 \pi \Phi}{N_\perp(\omega_\mathrm{L})} = \tilde \Phi(\omega_\mathrm{L})$. Although the longitudinal photon flux is not the correct physical quantity to characterize the CW laser we always relate it to a physical laser power output, $P_\mathrm{L}$.

Using the self-energy Eq.~\eqref{eq:Sigma_coherent_state} we are now in position to compute an equilibrium distribution function for the cavity mode. We are looking to write the lesser and greater cavity field propagator as $D^<(\omega) = -2\im f_\mathrm{cav}(\omega) A_\mathrm{cav}(\omega)$ and $D^>(\omega) = -2\im (f_\mathrm{cav}(\omega) + 1) A_\mathrm{cav}(\omega)$, where $A_\mathrm{cav}(\omega) = -2\Imag{D^\mathrm{r}(\omega)}$ is the cavity spectral function. 
If we neglect the effect of the polarization of the media on the distribution function of the cavity photons, we obtain using the Keldysh equation $D^< = D^r \Sigma^<_\mathrm{pump} D^a$. In the case of a single mode cavity field, this equation is algebraic and we can express the product $D^\mathrm{r}(\omega)\times D^a(\omega) = \mathcal{A}(\omega)/\Gamma^\mathrm{ph}$ where $\mathcal{A}(\omega) = -2\Imag{D^\mathrm{r}(\omega)}$ is the cavity spectral function and $\Gamma^\mathrm{ph} = -2\Imag{\Sigma^r_\mathrm{pump}(\omega)}$ is the photonic broadening function. Here, $\Gamma^\mathrm{ph} = \im \left[\Sigma_\mathrm{pump}^> - \Sigma_\mathrm{pump}^< \right]$.
The broadening function is thus $\Gamma^\mathrm{ph}(\omega) = +  \kappa \mathrm{sign}(\omega)$ and we can write the cavity propagators as 
\begin{align}
    D^<(\omega) &= \frac{-2\Imag{D^\mathrm{r}(\omega)}\Sigma^<_\mathrm{pump}}{\kappa \mathrm{sign}(\omega)} = +2\im f_\mathrm{cav}(\omega)\Imag{D^\mathrm{r}(\omega)} \nonumber \\
    D^>(\omega) &= \frac{-2\Imag{D^\mathrm{r}(\omega)}\Sigma^>_\mathrm{pump}}{\kappa \mathrm{sign}(\omega)} = +2\im \bar f_\mathrm{cav}(\omega)\Imag{D^\mathrm{r}(\omega)} ~,
\end{align}
where we defined the cavity distribution function as
\begin{align}
    f_\mathrm{cav}(\omega) &= \frac{\im\Sigma^<_\mathrm{pump}(\omega)}{\kappa \mathrm{sign}(\omega)} \nonumber \\
    &= \Theta(\omega) \tilde \Phi(\omega_\mathrm{L})  \delta(\omega - \omega_\mathrm{L}) - \Theta(-\omega)(\tilde \Phi(\omega_\mathrm{L})  \delta(\omega + \omega_\mathrm{L})  + 1) \nonumber \\
    \bar f_\mathrm{cav}(\omega) &= \frac{\im\Sigma^>_\mathrm{pump}(\omega)}{\kappa \mathrm{sign}(\omega)} \nonumber \\
    &= \Theta(\omega)(\tilde \Phi(\omega_\mathrm{L}) \delta(\omega - \omega_\mathrm{L})  + 1)  - \Theta(-\omega)\tilde \Phi(\omega_\mathrm{L}) \delta(\omega + \omega_\mathrm{L})~.
\end{align}

Using the same approximation as in Sec.~\ref{dark_current} for the polariton propagator and the pseudoequilibrium distribution function, it is possible to obtain analytical formula for the radiative self-energy under illumination,
\begin{align}
    \label{eq:2SB_radiative_illluminated_self_energy_analytic}
    \Sigma_{\mathrm{rad},n}^<(k,\omega) = +2\im  g_{12}^2f_\mathrm{QW}(\epsilon_{n'}(\vect{k}))\sum_\sigma &\Zsigtilde \Bigg\{\frac{\kappa \Zsigtilde  \Theta(\epsilon_{n'}(\vect{k}) - \omega)(\tilde \Phi(\omega_\mathrm{L}) \delta(\epsilon_{n'}(\vect{k}) - \omega - \omega_\mathrm{L})  + 1)}{(\epsilon_{n'}(\vect{k}) - \omega - \psigmatilde)^2 + (\kappa \Zsigtilde)^2} \nonumber \\
    &+ \frac{\kappa \Zsigtilde\Theta(\omega -\epsilon_{n'}(\vect{k}))\tilde \Phi(\omega_\mathrm{L})  \delta(\epsilon_{n'}(\vect{k}) - \omega + \omega_\mathrm{L})}{(\epsilon_{n'}(\vect{k}) - \omega + \psigmatilde)^2 + (\kappa \Zsigtilde)^2} \Bigg\} \nonumber \\
    \Sigma_{\mathrm{rad},n}^>(k,\omega) = -2\im g_{12}^2 \bar f_\mathrm{QW}(\epsilon_{n'}(\vect{k}))\sum_\sigma &\Zsigtilde \Bigg\{\frac{\kappa \Zsigtilde  \Theta(\epsilon_{n'}(\vect{k}) - \omega)\tilde \Phi(\omega_\mathrm{L}) \delta(\epsilon_{n'}(\vect{k}) - \omega - \omega_\mathrm{L})}{(\epsilon_{n'}(\vect{k}) - \omega - \psigmatilde)^2 + (\kappa \Zsigtilde)^2} \nonumber \\
    &+ \frac{\kappa \Zsigtilde\Theta(\omega -\epsilon_{n'}(\vect{k}))(\tilde \Phi(\omega_\mathrm{L}) \delta(\epsilon_{n'}(\vect{k}) - \omega + \omega_\mathrm{L}) + 1)}{(\epsilon_{n'}(\vect{k}) - \omega + \psigmatilde)^2 + (\kappa \Zsigtilde)^2} \Bigg\}
\end{align}
One can recognize in Eq.~\eqref{eq:2SB_radiative_illluminated_self_energy_analytic} the form of the light-matter interaction self-energy used to describe the photovoltaic effect \cite{Henrickson_2002, Aeberhard_2012} where the photonic propagator has been substituted with our polariton-pole approximation Eq.~\eqref{eq:polariton_pole_approx} and the equilibrium distribution function has been replaced by the pseudo-equilibrium one. 
In Eq.~\eqref{eq:2SB_radiative_illluminated_self_energy_analytic} the term proportional to ($\alpha_\mathrm{L}^2 + 1)$ corresponds to stimulated emission and spontaneous emission while the term proportional to $\alpha_\mathrm{L}^2$ corresponds to absorption. 

One could try now to use the above expression to express the energy shift under illumination, however the contour integration requires to convert the Dirac distribution into a Lorentzian, and the number of poles becomes inconvenient. We therefore chose to compute numerically the energy shift under illumination using the property $\Lambda_n(\vect{k}, \omega) = \frac{1}{2}\mathcal{H}\left[\Gamma_n(\vect{k}, \omega)\right]$ where $\mathcal{H}$ is the Hilbert transform.

% and we see that this quantity is, indeed, always zero. Thus we obtain a mathematically true current conservation. We call it here a \textit{pseudo-conservation} for two reasons: first, the self-energy that we obtained by inserting the pseudo-equilibrium distribution inside the bare Green's function cannot be obtained rigorously as a $\Phi$-derivable quantity in the sense of Baym. Secondly, the electron spectral function entering the current formula is still different from the one entering the radiative self-energy ; this lack of self-consistency will make us loose some phenomena, such as multi-photon absorption for instance. Nevertheless, the pseudo-conservation presented above as been used in literature to alleviate the heavy computational cost of SCBA approximation while retaining the physics of the problem \cite{}

% Create the reference section using BibTeX:

\bibliography{biblio.bib}